\documentclass[final,3p,12pt]{elsarticle}

\usepackage{amssymb}
\usepackage{amsmath}

\usepackage{graphicx}
\usepackage{subfig}
\usepackage{url}
\usepackage{amsfonts}
\usepackage{xcolor} 
\usepackage{tikz}
\usetikzlibrary{calc}  
\usepackage{caption}
\usetikzlibrary{arrows.meta}
\usepackage{booktabs}
\usepackage{multirow}
\journal{Engineering Applications of Artificial Intelligence}

\begin{document}

\begin{frontmatter}


\title{A framework for realisable data-driven active flow control using model predictive control applied to a simplified truck wake}

\author[INTA,UC3M]{Alberto Solera-Rico}
\ead{alberto.solera@alumnos.uc3m.es}

\author[INTA,UC3M]{Carlos Sanmiguel Vila}
\ead{csanmigu@ing.uc3m.es}

\author[UC3M]{Stefano Discetti\corref{cor1}}
\ead{sdiscett@ing.uc3m.es}

\cortext[cor1]{Corresponding author}

\affiliation[INTA]{organization={Sub-directorate general of aeronautical systems, Spanish National Institute for Aerospace Technology (INTA)},
            addressline={ctra. M-301, km. 10.5}, 
            city={San Martín de la Vega},
            postcode={28330}, 
            state={Madrid},
            country={Spain}}

\affiliation[UC3M]{organization={Universidad Carlos III de Madrid, ROR: https://ror.org/03ths8210, Departamento de Ingeniería Aeroespacial},
            addressline={Avenida de la Universidad, 30}, 
            city={Leganés},
            postcode={28911}, 
            state={Madrid},
            country={Spain}}

\begin{abstract}
We present  a data-driven active flow control framework designed for deployment  with few non-intrusive sensors.  The method builds upon Artificial Intelligence driven reduced-order predictive models based on Long-Short-Term Memory (LSTM) networks and efficient gradient-based Model Predictive Control (MPC). The model uses only surface-mounted pressure probes to infer the wake state, and is trained entirely offline on a dataset built with open-loop actuations, thus avoiding the complexities of online learning. Sparsification of the sensors needed for control from an initially large set is achieved using SHapley Additive exPlanations (SHAP). A parsimonious set of sensors is then deployed in closed-loop control with MPC. The framework is tested in numerical simulations of a two-dimensional truck model at Reynolds number 500, with pulsed-jet actuators placed in the rear of the truck to control the wake. The resulting LSTM-MPC achieved a drag reduction of 12.8\%.
\end{abstract}




\end{frontmatter}


\section{\label{sec:introduction}Introduction}

Active Flow Control (AFC) has emerged as a promising approach to reducing aerodynamic drag in ground and air vehicles. Manipulating separation and recovering base pressure with actuators such as steady and synthetic jets, plasma devices, and movable surfaces can produce meaningful performance gains, as surveyed in foundational reviews \cite{Greenblatt2000separationcontrol,Glezer2002reviewJets,Cattafesta2011reviewActuators}. For ground vehicles in particular, wind tunnel studies in squareback or Ahmed-type models demonstrate drag reduction by steady and pulsed blowing with realistic actuator layouts \cite{littlewood2012squareback,mcnally2015drag,cerutti2020van,amico2022deep, amico2024flow}. Moving from these demonstrations to practical, real-world systems, however, requires closed-loop real-time control implementations that are both robust and computationally efficient \cite{brunton2015closed}. To meet these demands, data-driven methods have become a leading strategy for developing predictive models \cite{Brunton2020mlFluids, regazzoni2024ldnets}, which can be leveraged in the optimisation and control process.


A prominent strategy that leverages such data-driven predictive models is Model Predictive Control (MPC). It offers an interesting framework for AFC thanks to its ability to include constraints and adapt to disturbances and model inaccuracies \cite{CamachoBordons2013_MPC, lee2011mpc}. MPC is based on a predictive model of the system to forecast its future evolution under control actions. The control is optimised over a receding horizon, while respecting hard or soft constraints on the actuation and the system state itself. The optimised control is applied only for a short time, and then the optimisation is repeated again with updated state information. MPC can use predictive models trained entirely offline from existing data of the system undergoing open-loop control. Furthermore, MPC provides significant flexibility by decoupling the predictive model from the control objective. Control objectives are defined in an explicit cost function, which can be dynamically modified at deployment time to prioritise different goals, such as balancing drag reduction against energy consumption or enforcing constraints on actuator limits, without having to retrain the underlying model. This contrasts with other methods, such as typical reinforcement learning formulations, where the control objective is implicitly embedded within the trained policy through the reward function, often requiring extensive retraining to adapt to new goals or constraints.

While this flexibility in the cost function is a key advantage, the effectiveness of such MPC strategies is still intrinsically related to the predictive power and computational efficiency of the underlying model. Pioneering early efforts in fluid mechanics focused on using an exact plant for MPC, with the aim of discovering control strategies to relaminarise a channel flow with blowing/suction \cite{bewley2001dns}. Recent studies have successfully implemented deep model predictive control using a deep neural network for the plant model \cite{morton2018deep}, and extended it for the case of limited sensors \cite{bieker2020deep}. Self-tuning MPC frameworks have also been proposed, with automatic hyperparameter optimisation during training  \cite{Marra2024Self}.

Underpinning many of these successful data-driven strategies is the hypothesis that the dynamical system to be modelled evolves on a low-dimensional attractor. Methods based on discovering governing equations from data \cite{brunton2016sindy, kaiser2018sindyMPC} or building deep-network predictors based on latent dynamics models have proven powerful in fluid flows. A low-dimensional compressed representation of the state of the system is learned, allowing highly efficient prediction and control \cite{PLDM, regazzoni2024ldnets, solera2024bvae, liu2025mbrl, fukagata2025compressing}. The addition of a physical properties decoder to condition the low-dimensional representation with physical variables has proven to be a fruitful approach in various related fields \cite{fukami2023grasping, fukami2025multisource}.

A natural next step is to embed these learned dynamics models directly into the MPC optimisation loop and exploit their differentiability for gradient-based control. Early formulations achieved this for linear systems with quadratic costs, either by differentiating through the fixed point of a Linear-Quadratic Regulator solver \cite{amos2018diffmpc} or by extending the framework to infinite-horizon settings with stability guarantees \cite{east2020infinite}. Relaxing the linearity assumption, subsequent work showed that neural state-space models trained from data can replace analytical plant descriptions within the MPC pipeline, with the control loss backpropagated through the closed-loop rollout to learn a parametric policy offline \cite{drgona2021dpc}. In parallel, the connection between residual network layers and Euler integration has been exploited to build surrogate plant models compatible with conventional nonlinear MPC solvers \cite{blaud2022resnet}. Our framework draws on both threads: the plant model is a nonlinear latent dynamics surrogate with a residual architecture, but rather than training an offline policy or relying on a generic nonlinear solver, we optimise the action sequence online at each control step by backpropagating directly through the unrolled model. This retains the flexibility of MPC, since the cost function can be modified at deployment without retraining, while remaining applicable to systems whose dynamics lack an analytical description.

Deploying such a controller also requires deciding which surface locations to instrument, and here a tension arises between classical and learning-based approaches. The dominant strategy in fluid mechanics applies QR pivoting to a truncated optimal basis from Proper Orthogonal Decomposition, selecting locations that maximise the volume of the measurement submatrix \cite{manohar2018data}; convex relaxation methods ~\cite{joshi2009sensor} provide near-optimal guarantees within the same linear framework. Both approaches, however, identify the sensors most informative for reconstructing a linear modal expansion, which need not coincide with those most relevant to a nonlinear encoder trained end-to-end for control. Ranking sensor importance by SHapley Additive exPlanations (SHAP) of the trained model, on the other hand, allows focusing on relevant information for the model predictive capability without requiring a separate data decomposition. Vishwasrao et al. ~\cite{vishwasrao2025diffsport} recently showed that SHAP-based placement produces spatially coherent sensor configurations that match or exceed QR-pivoting performance in urban flow reconstruction, particularly when the sensor budget is tight.

 Despite these individual advances, no existing framework addresses sensor selection, offline surrogate training, and computationally efficient gradient-based MPC within a single, deployment-oriented pipeline. Many prominent studies rely on intrusive probes placed directly in the wake \cite{Rabault2019, castellanos2022, Deda2024}, which provide a clean observation of the flow state but are often impractical for real-world applications. Furthermore, reliance on closed-loop online training remains a major barrier to straightforward deployment. Finally, the computational cost of the MPC optimisation loop itself can be a bottleneck for real-time implementation if the underlying model is not suitable for efficient gradient-based methods \cite{rawlings2017book}. These practical hurdles motivate the development of frameworks designed from the outset for experimental feasibility and computational tractability.

This study presents an end-to-end framework for developing a practical and computationally efficient AFC system that directly addresses these challenges. Our goal is to create a robust controller using a methodology that is based on realistic sensing and  informed by the practical constraints of potential experimental implementation. The key contributions are threefold:
1) We develop a predictive model that relies exclusively on nonintrusive, surface-mounted pressure sensors, inferring the wake state from its surface pressure footprint;
2) a feature attribution method \cite{SHAP} is used to analyse the model and select the most important sensors to train a more efficient encoder network;
3) we implement the entire predictive model in a deep learning framework that supports automatic differentiation. This allows MPC optimisation to be performed with highly efficient gradient-based methods, making real-time control computationally feasible \cite{amos2018diffmpc}.
Finally, we demonstrate the efficacy of this framework by deploying the controller in a 2D simulation environment of the wake of a truck, using a minimal set of just four sensors identified through an interpretable analysis to achieve drag reduction.

The paper is organised as follows. Section \ref{sec:methodology} provides a description of the methodology, including the data generation for the chosen test case, system modelling, sensor selection and control implementation. The analysis of the model and sensor selection, along with results of the control application are provided in \S~\ref{sec:results}. Finally, the conclusions are discussed in \S~\ref{sec:conclusion}.

\section{\label{sec:methodology}Methodology}

The methodology is designed with the aim of demonstrating the MPC framework and the sensor optimisation in a simulation environment, in which performances can be unambiguously assessed. The rationale of the method, nonetheless, is targeted to the practical application, i.e., parsimonious use of sensors and computationally slender predictive modelling. First, we generate a comprehensive dataset using random open-loop actuation sequences that serve as the basis for training our predictive model. This offline approach avoids the complex online training with the real system. Second, we train a model in the latent coordinate space to predict the aerodynamic forces from a history of surface pressure readings. Finally, we reduce the number of required sensors and train a lightweight ``slim'' encoder for its use in a closed-loop control system. This approach enables the use of model-based and data-driven control strategies that are  amenable to future experimental deployment.

\subsection{Flow configuration and data generation}

As a test bench for the framework, we use a high-fidelity Direct Numerical Simulation (DNS) of a simplified 2D truck model. DNS provides an accurate representation of flow dynamics, capturing all scale features of the wake with minimal numerical approximations.

Figure~\ref{fig:case} illustrates the flow configuration based on the horizontal mid plane geometry of the truck model from Ref.~\cite{GroundTransportationSystem}. The model consists of a rectangular bluff body with width $W=1$, length $L = 7.647W$, and rounded leading edges with radius $r=0.118W$. The inflow velocity is uniform with a magnitude of $U_\infty$, and the Reynolds number, defined as $Re = U_\infty W / \nu$, where $\nu$ represents the kinematic viscosity of the fluid, is set to $500$. The computational domain is rectangular, extending from $(-7W, 23W)$ in the streamwise ($x$) direction and $(-7.5W, 7.5W)$ in the transverse ($y$) direction, with the front of the bluff body placed at $x = 0$. The simulation domain is discretised with a hybrid mesh, structured near the wall and unstructured in the remaining region, spanning approximately $168,000$ cells. Time is non-dimensionalised using the convective time $t_c=W/U_\infty$. The DNS simulation is performed in OpenFOAM, using the Gym-preCICE \cite{gymprecice} wrapper for the \cite{preCICEv2} coupling library and the OpenFOAM adapter \cite{OpenFOAMpreCICE} to couple the simulation with the controller.

\begin{figure}[htbp]
\centering
\begin{tikzpicture}[
    scale=1.3, 
    annot/.style={-{Stealth[length=2mm, width=1.5mm]}},
    every node/.style={font=\sffamily\footnotesize},
    point/.style={circle, fill=blue, inner sep=1pt} 
]

\def\rectwidth{7.65}
\def\rectheight{1}
\def\radius{0.118}

\def\jetwidth{0.05}
\def\jetlength{0.3} 

\draw[thick] 
    (\rectwidth, \rectheight/2) 
    -- (\radius, \rectheight/2) 
    arc(90:180:\radius)
    -- (0, -\rectheight/2 + \radius)
    arc(180:270:\radius)
    -- (\rectwidth, -\rectheight/2)
    -- cycle;

\foreach \x/\y in {
    0.301634/0.500000, 0.485268/0.500000, 0.668902/0.500000, 
    0.852537/0.500000, 1.036171/0.500000, 1.219805/0.500000, 
    1.403439/0.500000, 1.587073/0.500000, 1.770707/0.500000, 
    1.954341/0.500000, 2.137976/0.500000, 2.321610/0.500000, 
    2.505244/0.500000, 2.688878/0.500000, 2.872512/0.500000, 
    3.056146/0.500000, 3.239780/0.500000, 3.423415/0.500000, 
    3.607049/0.500000, 3.790683/0.500000, 3.974317/0.500000, 
    4.157951/0.500000, 4.341585/0.500000, 4.525220/0.500000, 
    4.708854/0.500000, 4.892488/0.500000, 5.076122/0.500000, 
    5.259756/0.500000, 5.443390/0.500000, 5.627024/0.500000, 
    5.810659/0.500000, 5.994293/0.500000, 6.177927/0.500000, 
    6.361561/0.500000, 6.545195/0.500000, 6.728829/0.500000, 
    6.912463/0.500000, 7.096098/0.500000, 7.279732/0.500000, 
    7.463366/0.500000, 0.301634/-0.500000, 0.485268/-0.500000, 
    0.668902/-0.500000, 0.852537/-0.500000, 1.036171/-0.500000, 
    1.219805/-0.500000, 1.403439/-0.500000, 1.587073/-0.500000, 
    1.770707/-0.500000, 1.954341/-0.500000, 2.137976/-0.500000, 
    2.321610/-0.500000, 2.505244/-0.500000, 2.688878/-0.500000, 
    2.872512/-0.500000, 3.056146/-0.500000, 3.239780/-0.500000, 
    3.423415/-0.500000, 3.607049/-0.500000, 3.790683/-0.500000, 
    3.974317/-0.500000, 4.157951/-0.500000, 4.341585/-0.500000, 
    4.525220/-0.500000, 4.708854/-0.500000, 4.892488/-0.500000, 
    5.076122/-0.500000, 5.259756/-0.500000, 5.443390/-0.500000, 
    5.627024/-0.500000, 5.810659/-0.500000, 5.994293/-0.500000, 
    6.177927/-0.500000, 6.361561/-0.500000, 6.545195/-0.500000, 
    6.728829/-0.500000, 6.912463/-0.500000, 7.096098/-0.500000, 
    7.279732/-0.500000, 7.463366/-0.500000, 7.647000/0.368182, 
    7.647000/0.286364, 7.647000/0.204555, 7.647000/0.122727, 
    7.647000/0.040909, 7.647000/-0.040909, 7.647000/-0.122727, 
    7.647000/-0.204545, 7.647000/-0.286364, 7.647000/-0.368182
} {
    \node[point] at (\x,\y) {};
}

\fill[red, opacity=0.8]
    (\rectwidth, \rectheight/2) 
    .. controls (\rectwidth + \jetlength, \rectheight/2) and (\rectwidth + \jetlength, \rectheight/2 - \jetwidth) .. (\rectwidth, \rectheight/2 - \jetwidth)
    -- cycle;

\fill[red, opacity=0.8]
    (\rectwidth, -\rectheight/2)
    .. controls (\rectwidth - \jetlength, -\rectheight/2) and (\rectwidth - \jetlength, -\rectheight/2 + \jetwidth) .. (\rectwidth, -\rectheight/2 + \jetwidth)
    -- cycle;
    
\draw (0,-\rectheight/2) -- ++(0,-0.5); 
\draw (\rectwidth,-\rectheight/2) -- ++(0,-0.5); 
\draw[{Latex[length=1.5mm]}-{Latex[length=1.5mm]}] (0,-0.8) -- node[below] {$L=7.65W$} (\rectwidth,-0.8);

\draw (0, \rectheight/2) -- ++(-0.5, 0); 
\draw (0, -\rectheight/2) -- ++(-0.5, 0); 
\draw[{Latex[length=1.5mm]}-{Latex[length=1.5mm]}] 
    (-0.3, -\rectheight/2) -- node[left] {$W$} (-0.3, \rectheight/2);

\coordinate (R_annot_start) at (-\rectheight*.25, \rectheight/2 + 0.4);
\draw[annot] (R_annot_start) -- (0, \rectheight/2);
\node[left=2pt] at (R_annot_start) {$r=0.118W$};

\coordinate (Wj_annot_start) at (\rectwidth-\rectheight*.25, \rectheight/2 + 0.4);
\draw[annot] (Wj_annot_start) -- (\rectwidth, \rectheight/2);
\node[left=1pt] at (Wj_annot_start) {$w_{jet}= 0.05W$};

\draw[-{Stealth[length=4mm, width=3mm]}, line width=1.25pt] (-1.8*\rectheight, 0) -- (-\rectheight, 0);
\node[above=2pt] at (-\rectheight*1.5, 0) {$U_{\infty}$};

\end{tikzpicture}
\caption{Schematic of the flow configuration. Sensor locations depicted in blue and zero-net-mass flow jets schematised in red.}
\label{fig:case}
\end{figure}

\begin{figure}[htbp]
  \centering
  \def\mainwidth{0.60\textwidth}   
  \def\detailwidth{0.3\textwidth} 
  \def\xsep{1cm}                 

  \begin{tikzpicture}[>=stealth]
    \node[anchor=south west, inner sep=0] (main)
      at (0,0) {\includegraphics[width=\mainwidth]{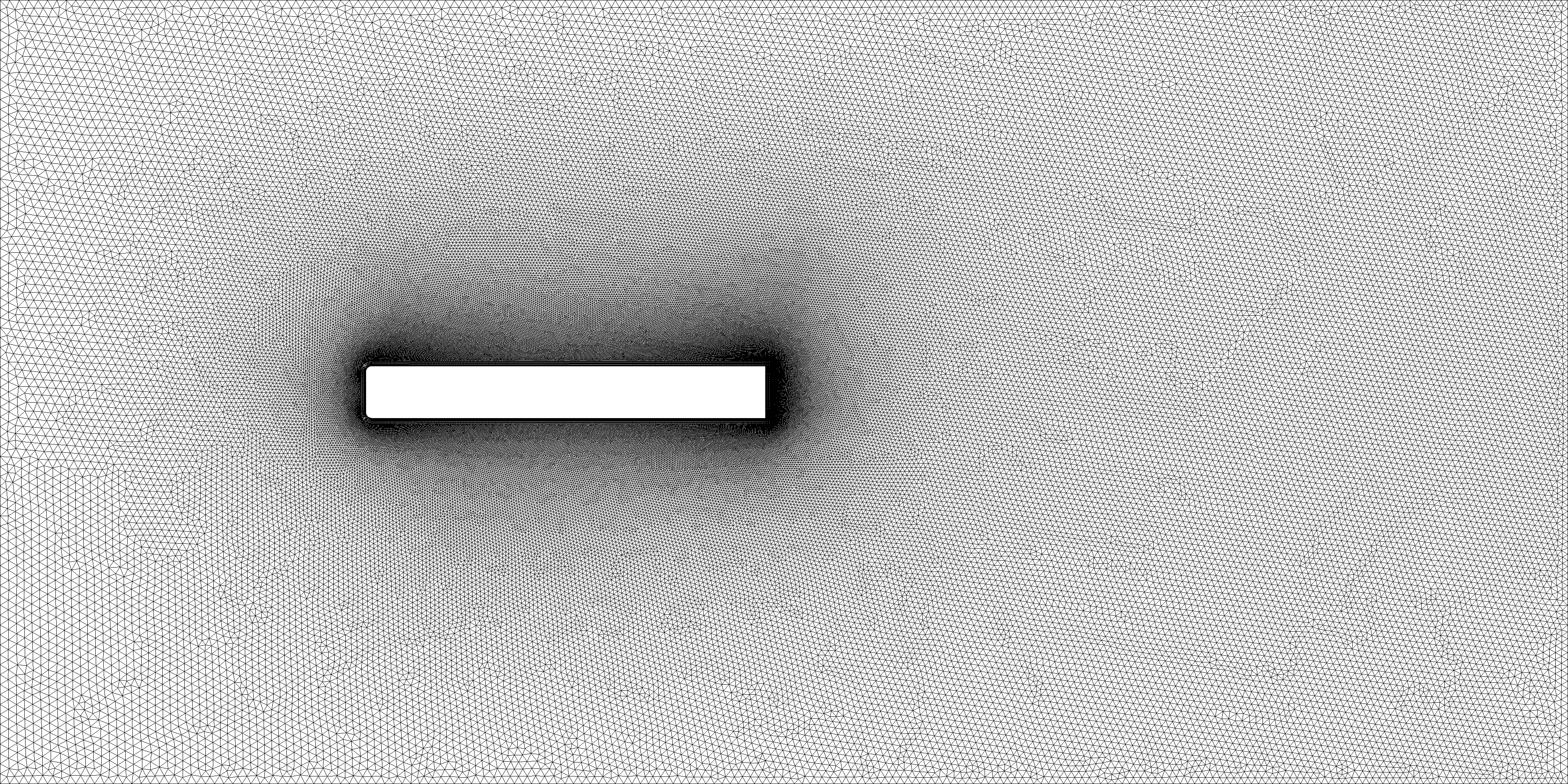}};

    \node[anchor=west, inner sep=0] (detail)
      at ($(main.east) + (\xsep,0)$) {\includegraphics[width=\detailwidth]{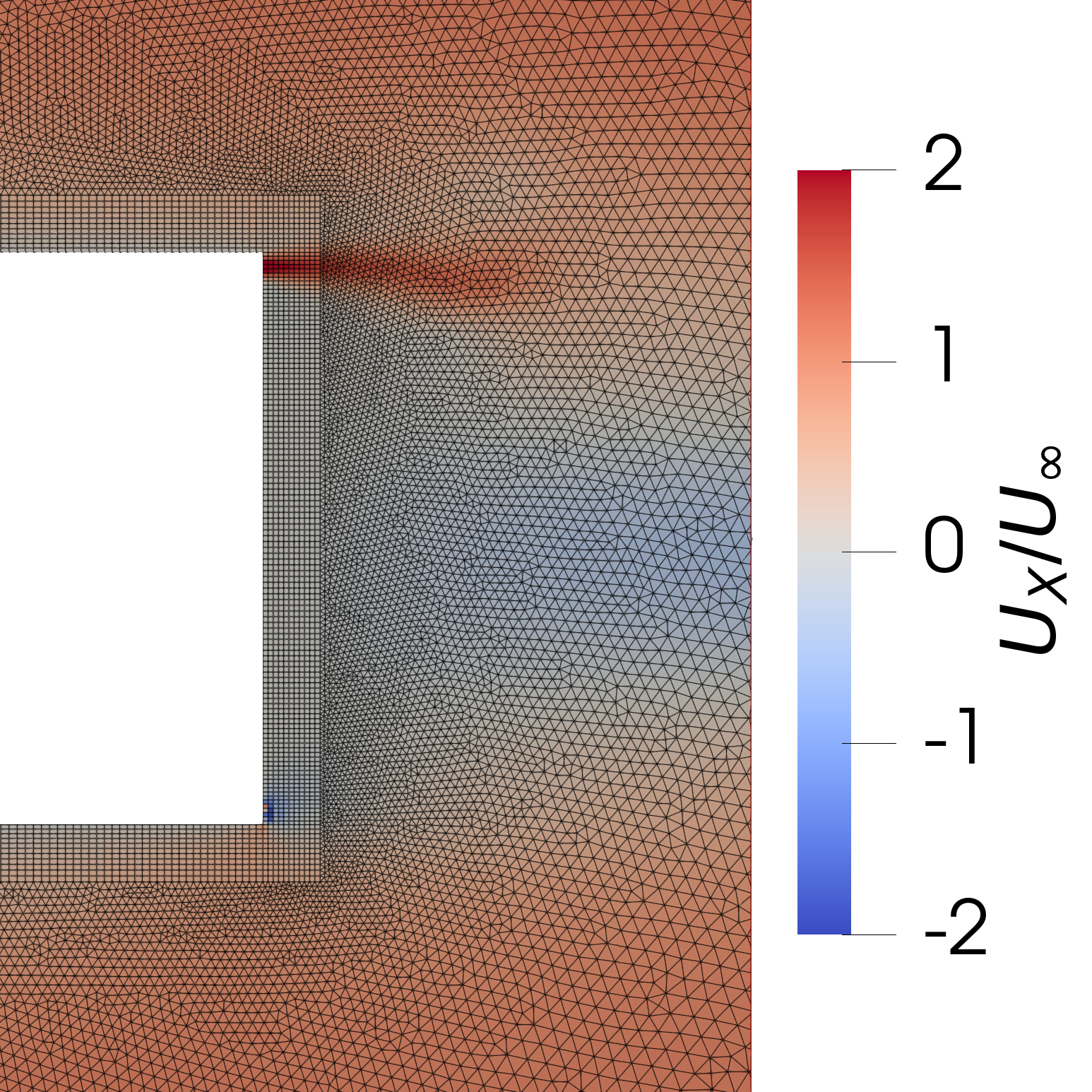}};

    \draw[line width=1pt, color=blue, rounded corners=1pt] (detail.north west) rectangle (detail.south east);


    \begin{scope}[shift={(main.south west)}, x={(main.south east)}, y={(main.north west)}]
      \coordinate (cropSW) at (0.47,0.42);  
      \coordinate (cropNE) at (0.53,0.58);  
      \coordinate (cropNW) at (cropSW |- cropNE); 
      \coordinate (cropSE) at (cropNE |- cropSW); 

      \draw[line width=1pt, color=blue, rounded corners=1pt] (cropSW) rectangle (cropNE);

    \end{scope}

    \draw[blue, dashed] (cropSW) -- (detail.south west);
    \draw[blue, dashed] (cropNW) -- (detail.north west);
  \end{tikzpicture}
    \caption{DNS mesh and detail of the flow near the rear of the truck, with the jets actuating at maximum suction/blowing intensity.}
    \label{fig:case_jets}
\end{figure}

The control is achieved by two opposite-flow jets located on the sides of the vehicle base, with zero-net mass flow. This configuration is similar to previous experimental studies \cite{barros2016coanda, cerutti2020van}. The jets have a width of $w_{jet}=0.05W$ and produce parabolic velocity profiles with a maximum mean velocity of $1.5U_\infty$. The resulting dataset consists of a time series of pressure sensor readings, control intensity, and force data, with a time step of $\Delta t = t_c/5 = W/{(5U_\infty)}$. The global mesh and a sample of the flow around the jets are shown in Figure~\ref{fig:case_jets}.

The aerodynamic forces in the body are expressed in terms of the dimensionless drag and lift coefficients, $C_d$ and $C_l$, which are computed directly from the OpenFOAM simulations. These coefficients are defined as
\begin{equation}
C_d = \frac{F_x}{\frac{1}{2} \rho U_\infty^2 A}, \quad
C_l = \frac{F_y}{\frac{1}{2} \rho U_\infty^2 A},
\end{equation}
where $F_x$ and $F_y$ are the streamwise and transverse forces acting on the body, $\rho$ is the fluid density and $A$ is the reference area of the bluff body. For the present 2D configuration, the reference area is taken as the width of the body $W$ multiplied by a unit depth. These coefficients integrate the pressure and viscous stress distributions over the body surface.


The training data was generated using forcing with an open-loop control. The control signal was specifically designed to excite the expected sensitive frequency range around the natural shedding frequency $f_{sh}\approx0.2$.
The signal is a synthesised waveform characterised by simultaneous frequency and amplitude modulation, constructed as follows:

\begin{enumerate}
    \item \textbf{Carrier wave:} A base sinusoidal wave with a fundamental frequency of $f_{base}=0.2$, corresponding to the natural shedding frequency ($f_{sh}$) of the wake, serves as the carrier.

    \item \textbf{Frequency Modulation (FM):} To explore a range of temporal scales, the carrier frequency is modulated by a randomly varying signal. This is achieved by generating white noise and applying a second-order low-pass Butterworth filter with a cut-off frequency of $0.05$ ($f_{sh}/4$). The resulting filtered noise, $\eta(t)$, is normalised to $[-1, 1]$ and modulates the instantaneous frequency according to $f(t) = f_{base} + k_{fm} \cdot \eta(t)$, with a modulation index of $k_{fm}=0.25$. This produces a smooth, wandering frequency that explores a range up to a maximum of 0.45.

    \item \textbf{Amplitude Modulation (AM):} To ensure that the model learns the response of the system to the varying actuation power, the amplitude of the FM signal is modulated by a slow sinusoidal envelope with a frequency of $0.005$ ($f_{sh}/40$) and an amplitude depth of 45\%. This causes the overall magnitude of the actuation to vary smoothly between low- and high-power regimes.
\end{enumerate}

\begin{figure}[htbp]
    \centering
    \includegraphics[width=\textwidth]{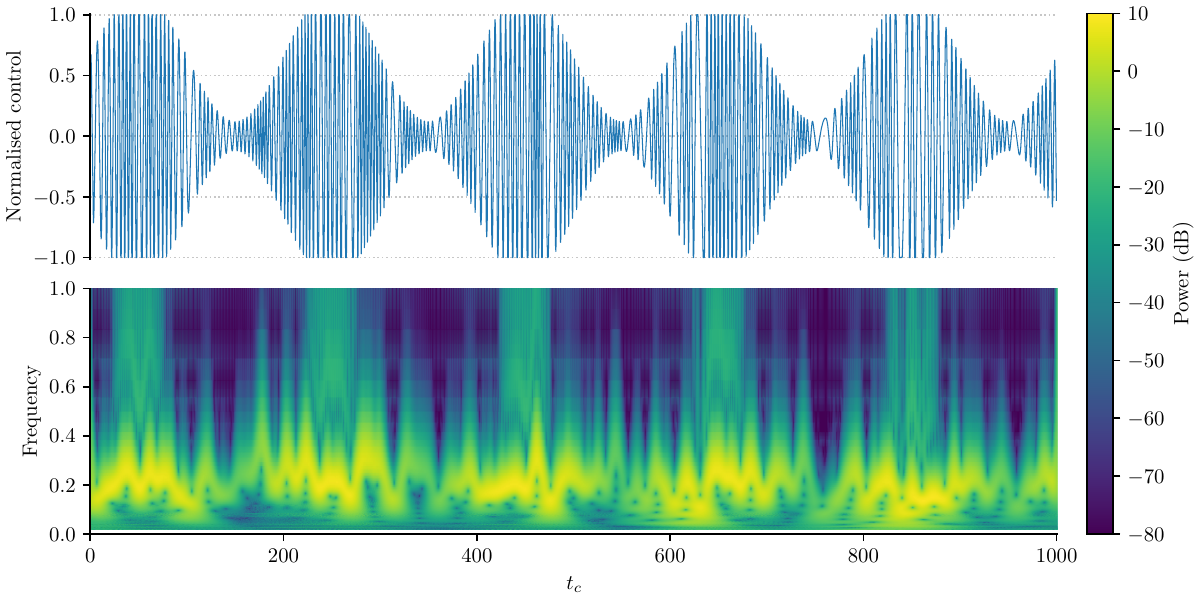}
    \caption{Sample portion of the modulated control signal applied during training data generation. Top: normalised signal. Bottom: Wavelet spectrogram of the same signal to show the frequency content over time.}
    \label{fig:control}
\end{figure}

The final synthesised signal, shown in Figure~\ref{fig:control}, is clipped to a normalised range of $[-1, 1]$ and then scaled to match the physical action limits of the synthetic jet, corresponding to a maximum dimensionless flow rate of $\pm 0.075 U_\infty W$.

During data generation simulation, the precomputed forcing signal was sampled at a control frequency of $f_s=5 t_c^{-1}$ ($\Delta t = 0.2 t_c$). The jet flow is linearly interpolated between these control updates to avoid discontinuities. This sequence spans $50,000$ time steps. A shorter $2,500$-step sequence of uncontrolled dynamics is also added to the training dataset to improve generalisation. The simulation records the resulting wall pressure data from the $N_s=90$ sensor probes and the integrated aerodynamic forces ($C_d$, $C_l$). The sensors are evenly spaced with $40$ on each side of the vehicle and $10$ on the back, as shown in Figure~\ref{fig:case}. The number of sensors is arbitrarily selected to provide a reasonable coverage of the pressure coefficient distribution on the lateral walls and the rear of the model.

The final dataset spans $52,500$ time steps. This data was then augmented with the symmetric state and actuation during training to enhance generalisation and symmetry of the model.

For validation and testing, a separate $10,000$ time steps dataset was used, generated with a linear chirp signal actuation. This chirp signal sweeps frequencies $f_{chirp}$ from $0.1$ to $0.4$. This frequency range is entirely contained within the frequency domain explored by the modulated training signal, which reached a maximum of $0.45$ as seen in Figure~\ref{fig:control}. This ensures that the test evaluates the ability of the model to accurately interpolate within its learnt dynamic range, rather than requiring it to extrapolate to unseen frequencies. The chirp signal also produces a small region of wake stabilisation, providing a valuable test case for the predictive capabilities of the model, as seen in Figure~\ref{fig:test_forces}.
The robustness of the model to out-of-distribution actuation frequencies and sensor noise is analysed in \ref{app:ood}.

\subsection{Latent dynamics model}

\begin{figure}[htbp]
    \includegraphics[width=\textwidth]{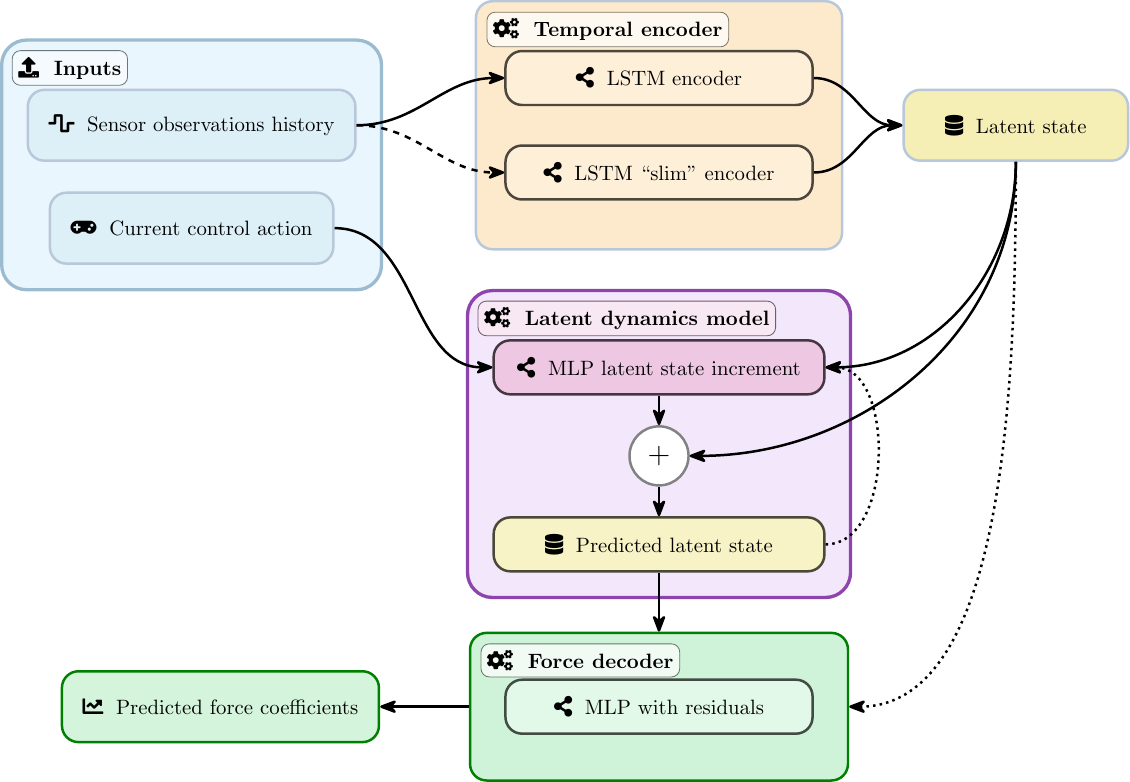}
    \caption{Schematic of the model architecture. A history of sensor observations $\mathbf{S}_t$ is mapped by the temporal encoder to a latent state $\mathbf{z}_t$. The action-aware dynamics model propagates this state forward to $\hat{\mathbf{z}}_{t+1}$ using the current control action $\mathbf{a}_t$. The force decoder maps the latent state $\mathbf{z}_t$ to the predicted aerodynamic force coefficients $\hat{\mathbf{C}}_t$.}
    \label{fig:pldm_architecture}
\end{figure}

The proposed method is based on mapping sensor observations to a low-dimensional, learnt latent space, thus avoiding direct prediction of the high-dimensional full state. Within this compressed space, a more compact predictive model for the state evolution is inferred. The architecture, shown schematically in Fig.~\ref{fig:pldm_architecture}, is composed of three end-to-end trained neural network modules: a temporal encoder, which performs dimensionality reduction; an action-aware dynamics model, which serves as the predictive engine in the latent space; and a force decoder, which translates the latent state into physical quantities of interest such as force coefficients. By training the models simultaneously, the resulting latent space is optimised to be informative for prediction and correlated with the aerodynamic forces, as shown in previous studies \cite{regazzoni2024ldnets, fukagata2025compressing}.

\subsubsection{Temporal encoder}
The first component, the temporal encoder, is responsible for feature extraction and dimensionality reduction. Its purpose is to distil the rich spatio-temporal information contained in the high-dimensional history of sensor measurements into a compact and generalisable latent state vector.

The input to the encoder at time $t$ is a sequence of the last $L=32$ sensor observations, denoted as $\mathbf{S}_t = \{\mathbf{s}_{t-L+1}, ..., \mathbf{s}_t\}$, where each $\mathbf{s} \in \mathbb{R}^{N_s}$ and $N_s$ is the number of sensors. The output is a single latent state vector $\mathbf{z}_t \in \mathbb{R}^{N_z}$, where the latent dimension is $N_z=8 \ll N_s \times L$. The latent dimension was selected by increasing it until the accuracy of the predictions made by the complete model reached a plateau. Increasing the latent dimension further tends to induce overfitting, whereas reducing it constrains the ability of the model to capture the dynamics of the system.

Given the sequential nature of the input, the encoder has been implemented using a Long-Short-Term Memory (LSTM) network \cite{Hochreiter1997lstm}. The LSTM processes the entire input sequence, and the hidden state from the final time step is passed through a Multi-Layer Perceptron (MLP) to produce the latent vector. This process is represented by the function $f_{\text{enc}}$:
\begin{equation}
\mathbf{z}_t = f_{\text{enc}}(\mathbf{S}_t; \theta_{\text{enc}})
\end{equation}
where $\theta_{\text{enc}}$ represents the trainable weights of the encoder network.

\subsubsection{Latent dynamics model}
The latent dynamics model is the predictive engine of the model. It functions as a discrete-time state transition model within the learnt latent space, forecasting the evolution of the system one step into the future based on its current state and the applied control action. It takes as input the current latent state $\mathbf{z}_t$ and the current control action $\mathbf{a}_t \in \mathbb{R}^{N_a}$. The output is the latent state predicted at the next time step, $\hat{\mathbf{z}}_{t+1}$.

The model first concatenates the current state $\mathbf{z}_t$ and the current action $\mathbf{a}_t$. The resulting vector passes through an MLP, which is trained to predict the change in the latent state $\Delta \mathbf{z}_t$. We adopt a residual connection, a common practice that stabilises training for dynamics models \cite{blaud2022resnet}, so that the final prediction is the sum of the current state and the predicted change:
\begin{equation}
\hat{\mathbf{z}}_{t+1} = \mathbf{z}_t + f_{\text{dyn}}(\mathbf{z}_t, \mathbf{a}_t; \theta_{\text{dyn}})
\end{equation}
where $f_{\text{dyn}}$ represents the latent dynamics network with trainable weights $\theta_{\text{dyn}}$.

\subsubsection{Force decoder}
The force decoder provides the link from the abstract latent space back to the physical quantities required for control and evaluation. Its function is to provide an instantaneous estimate of the aerodynamic force coefficients from a given latent state vector.

The decoder takes a latent state vector $\mathbf{z}_t \in \mathbb{R}^{N_z}$ as input and outputs the predicted aerodynamic force coefficients, $\hat{\mathbf{C}}_t = [\hat{C}_d, \hat{C}_l]_t \in \mathbb{R}^2$.

The decoder is implemented as an MLP with residual connections (ResNet-style architecture \cite{he2016deep}). This design facilitates the training of deeper networks as it reduces the problem of vanishing gradients during training. Although the decoder is composed only of a few layers, during training, the gradient must flow through the three models, which justifies the use of residual connections. The mapping is given by:
\begin{equation}
\hat{\mathbf{C}}_t = f_{\text{dec}}(\mathbf{z}_t; \theta_{\text{dec}})
\end{equation}
where $\theta_{\text{dec}}$ are the trainable weights of the decoder. This component is trained concurrently with the rest of the model, ensuring that the latent space becomes structured in a way that is not only predictable in time, but also informative about the instantaneous aerodynamic forces. This latent-space conditioning was previously explored in Ref.~\cite{fukami2023grasping} where it was proven to enhance the expressivity of the latent-space manifold with a coherent geometric structure.

\subsubsection{Training procedure}

First, the datasets are concatenated and the variables are standardised to zero mean and unit variance, with scaling parameters computed from the entire training dataset. The test dataset is scaled using these same parameters.

The three modules of the neural network (the encoder, the dynamics model, and the force decoder) are trained simultaneously in an end-to-end manner. This joint optimisation strategy ensures that the learnt latent space is not only predictable over time but also contains the necessary information for an accurate, instantaneous force estimation. Training is guided by a composite loss function $\mathcal{L}$, which combines objectives for latent state prediction, force prediction, and latent space regularisation:
\begin{equation}
\label{eq:loss_function}
\mathcal{L} = \mathcal{L}_{\text{pred}} + \lambda_{\text{force}} \mathcal{L}_{\text{force}} + 
\underbrace{\lambda_{\text{var}} \mathcal{L}_{\text{var}} + \lambda_{\text{cov}} \mathcal{L}_{\text{cov}}}_{\text{VICReg}}
\end{equation}
where the weighting hyperparameters are set to $\lambda_{\text{force}} = 0.25$, $\lambda_{\text{var}} = 0.02$, and $\lambda_{\text{cov}} = 0.02$.

The latent state prediction loss in (\ref{eq:loss_function}), $\mathcal{L}_{\text{pred}}$, penalises the discrepancy between the predicted latent state and the ground truth. To enhance long-term predictive stability, the loss is computed on an unrolled multistep prediction horizon. Starting from an initial state $\mathbf{z}_t$, the dynamics model is applied recursively for $k=5$ steps to generate a sequence of predictions $\{\hat{\mathbf{z}}_{t+1}, ..., \hat{\mathbf{z}}_{t+k}\}$. The ground truth sequence $\{\mathbf{z}_{t+1}, ..., \mathbf{z}_{t+k}\}$ is obtained by passing the corresponding true sensor histories through the encoder. The loss is $L_2$ norm computed over all steps of the sequence:
\begin{equation}
\mathcal{L}_{\text{pred}} = \frac{1}{k} \sum_{i=1}^{k} \left\| \hat{\mathbf{z}}_{t+i} - \mathbf{z}_{t+i} \right\|^2_2
\end{equation}

The force prediction loss in (\ref{eq:loss_function}), $\mathcal{L}_{\text{force}}$, ensures that the latent space remains physically meaningful by enforcing an accurate mapping of any latent state to its corresponding aerodynamic forces. At each step of the recursive prediction, the force decoder estimates the aerodynamic coefficients from the predicted latent state. This model uses dropout with a probability value $0.25$ between layers to mitigate overfitting. This loss term measures the difference between the predicted forces $\hat{\mathbf{C}}_{t+i}$ and the true forces from the dataset $\mathbf{C}_{t+i}$. We employ a $\text{Smooth}_{L_1}$ loss, which is less sensitive to outliers than $L_2$. It behaves like an $L_2$ loss for small errors and an $L_1$ loss for large errors, defined as:
\begin{equation}
\text{Smooth}_{L_1}(x) = 
\begin{cases} 
0.5 x^2 & \text{if } |x| < 1 \\
|x| - 0.5 & \text{otherwise}
\end{cases}
\end{equation}
where $x$ is the element-wise error. The total force loss is the average over the prediction horizon:
\begin{equation}
\mathcal{L}_{\text{force}} = \frac{1}{k} \sum_{i=1}^{k} \text{Smooth}_{L_1}(\hat{\mathbf{C}}_{t+i} - \mathbf{C}_{t+i})
\end{equation}

Finally, to prevent informational collapse and encourage a well-structured latent space, we introduce a regularisation loss inspired by the Variance-Invariance-Covariance (VICReg) methodology \cite{bardes2022vicreg}. This loss is composed of two parts that act on a batch of latent state predictions $\mathbf{Z} \in \mathbb{R}^{B \times D}$, where $B$ is the batch size and $D$ is the latent dimension.
\begin{itemize}
    \item The variance loss in (\ref{eq:loss_function}), $\mathcal{L}_{\text{var}}$, encourages the standard deviation of each latent dimension on the batch to be close to the unit. This prevents the representations from collapsing into a subspace.
    \begin{equation}
        \mathcal{L}_{\text{var}} = \frac{1}{D} \sum_{j=1}^{D} \max(0, 1 - \sqrt{\text{Var}(\mathbf{Z}_{:j}) + \epsilon})
    \end{equation}
    where $\epsilon$ is a small positive constant for numerical stability.
    \item The covariance loss in (\ref{eq:loss_function}), $\mathcal{L}_{\text{cov}}$, pushes the off-diagonal elements of the covariance matrix of the latent variables towards zero, decorrelating the latent dimensions.
    \begin{equation}
        \mathcal{L}_{\text{cov}} = \frac{1}{D(D-1)} \sum_{i \neq j} \left( \text{Cov}(\mathbf{Z}_{:i},\mathbf{Z}_{:j}) \right)^2
    \end{equation}
\end{itemize}

All trainable parameters in the three modules ($\theta_{\text{enc}}$, $\theta_{\text{dyn}}$, and $\theta_{\text{dec}}$) are updated jointly using the Adam optimiser \cite{adam2014method}, which minimises the total loss function $\mathcal{L}$ given in (\ref{eq:loss_function}) by backpropagating through the unrolled computation graph. The batch size is $B=512$, the learning rate is 0.001 and the training spans 300 epochs. The encoder processes a lookback window of $L = 32$ time steps with $N_s + 1 = 91$ input features (90 pressure sensors and 1 control signal) and maps them to an $N_z = 8$ dimensional latent space. The full architecture of each module is detailed in Table~\ref{tab:architecture}.


\begin{table}[t]
\centering
\caption{Neural network architecture for each module. $N_s$ denotes the number of sensors and $L$ the lookback window length.}
\label{tab:architecture}
\small
\begin{tabular}{@{}p{2.8cm}p{3cm}p{3.8cm}p{5.5cm}@{}}
\toprule
Module & Layer sequence & Dimensions & Details \\
\midrule
\multirow{2}{*}[-0.5em]{Temporal encoder}
  & LSTM (1 layer) & $(L \times (N_s+1)) \to 64$ & LayerNorm on output \\
  & MLP (2 layers) & $64 \to 64 \to 8$                   & ELU \\
\midrule
Latent dynamics
  & MLP (3 layers) & $(8+1) \to 256 \to 256 \to 8$ & LayerNorm, ELU, residual ($\Delta\mathbf{z}$) \\
\midrule
\multirow{3}{*}[-0.5em]{Force decoder}
  & Linear projection & $8 \to 128$   &\\
  & 2 residual blocks & $128 \to 128$ & LayerNorm, ELU, Dropout ($p\!=\!0.25$) \\
  & Linear output     & $128 \to 2$   & $C_d$, $C_l$ \\
\bottomrule
\end{tabular}
\end{table}

\subsection{Interpretable sensor selection}
A key challenge in developing practical flow control systems is identifying a minimal and realisable but effective set of sensors. Although the initial model is trained offline on a dense array of $90$ sensors to capture as much information as possible, deploying such a system would be costly and inefficient. We therefore implement a two-phase methodology to systematically reduce the number of sensors by ranking their importance and then training a new, computationally lightweight ``slim'' encoder that operates only on the most informative sensor subset. In this way, the full sensor stack needs only to be used in the initial open-loop dataset recording, while time-critical closed-loop implementation would rely on a small subset of sensors that can be more easily implemented.

\subsubsection{SHAP-based ranking}
To classify the sensors according to their importance, we interpret the trained temporal encoder using SHAP \cite{SHAP}, a framework from cooperative game theory \cite{shapley1953} to explain the output of the machine learning model. The Shapley value of a feature quantifies its average marginal contribution to a prediction in all possible combinations of features, providing a theoretically sound measure of importance. Since calculating exact Shapley values is computationally prohibitive, we employ the GradientExplainer algorithm from the SHAP library, which implements \emph{expected gradients}, an extension of \emph{integrated gradients} (IG) for differentiable models \cite{sundararajan2017axiomatic}. This method provides an efficient approximation for deep learning models by integrating the output gradients of the model with respect to the input features, using a distribution of background samples as a reference for a typical input. The process therefore requires two datasets: a set of background samples ($100$ in our case) to define the baseline distribution, and a set of explanation samples ($500$ in our case) for which the feature importances are calculated.

The explainer produces SHAP values that quantify the contribution of each of the $90$ sensors in each of the $32$ time steps in the history window to each of the dimensions in the output latent state vector $\mathbf{z}_t$. To obtain a single, robust importance score for each sensor, we aggregate these values. First, we take the absolute SHAP value to measure the magnitude of the contribution, regardless of its sign. These magnitudes are then summed across all latent dimensions and subsequently averaged over both the explanation samples and the history window. This yields a single importance score for each of the $90$ sensors.

This \textit{a posteriori} methodology offers various advantages over embedded selection techniques, such as those that promote input sparsity using $L_1$ regularisation or reinforcement learning in an input gate layer \cite{Deda2024, paris2021rlselection}. In those methods, the final number of active sensors is an indirect outcome of a sparsity hyperparameter, often requiring multiple training runs to achieve a specific target number of sensors. In contrast, our approach decouples sensor selection entirely from model training. By analysing a single fully-trained encoder, we generate an explicit importance ranking for all sensors. This provides complete flexibility, as the desired number of sensors becomes a design choice that can be made after the analysis. This not only simplifies the hyperparameter tuning process by removing the sparsity penalty term, but also provides a more direct and efficient path to designing sensor-reduced models tailored to specific needs or limitations.

To verify the stability of this ranking, the analysis was repeated ten times with independently drawn background and explanation samples; the top-$k$ sensor sets ($k \in \{2, 4, 6, 8\}$) were identical across all runs, confirming that the ranking is not an artefact of sample selection.


\subsubsection{Knowledge distillation for ``slim'' encoder}
Once the top \textit{N} most informative sensors are identified, a new, computationally efficient ``slim'' encoder is trained. This is achieved through a process of knowledge distillation, where the original 90-sensor encoder acts as a ``teacher'' network to train a smaller ``student'' network. The ``slim'' encoder shares the same architecture as the original temporal encoder, but its input dimension is reduced from $90$ to the selected \textit{N} sensors.

The objective is to distil the knowledge of the teacher model into the student, enabling the ``slim'' encoder to reproduce the latent space structure of the original model using only a fraction of the sensor input. During this process, the weights of the teacher encoder are frozen. For each batch of training data, the full $90$-sensor history $\mathbf{S}_t$ is passed through the teacher to generate a target latent vector $\mathbf{z}_{t, \text{target}}$. Simultaneously, the corresponding subset of the data from the \textit{N} selected sensors, $\mathbf{S}_{t, \text{slim}}$, is fed into the ``slim'' trainable encoder to produce a prediction, $\mathbf{z}_{t, \text{slim}}$. The student network is then optimised by minimising the $L_2$ norm between its output and the target vector of the teacher.
\begin{equation}
\mathcal{L}_{\text{distil}} = \frac{1}{B} \sum_{i=1}^{B} \left\| \mathbf{z}_{t, \text{target}}^{(i)} - \mathbf{z}_{t, \text{slim}}^{(i)} \right\|^2_2
\label{eq:distill_loss}
\end{equation}
where $B$ is the number of samples in the batch.

This procedure creates a lightweight encoder that requires significantly fewer sensors. Crucially, it can be used interchangeably with the original one for downstream tasks without retraining the latent dynamics model or the force decoder. This decoupling greatly reduces the physical and computational requirements for the potential deployment of experimental control systems.

\subsection{Latent-space model predictive control}
With the trained and validated model, we can formulate a closed-loop control strategy to minimise aerodynamic drag. MPC is a natural framework for this task, as it leverages the predictive capabilities of the model to make optimal decisions over a finite time horizon. A key advantage of our approach is that the entire optimisation process occurs within the computationally efficient, differentiable, low-dimensional latent space, which makes it suitable for real-time applications.

At each control step $t$, the controller uses the history of the last $L$ sensor observations $\mathbf{S}_t$ and the encoder to establish the initial state. The objective is to find an optimal sequence of future control actions, $\mathbf{U}^*_t = \{\mathbf{a}^*_t, \dots, \mathbf{a}^*_{t+H-1}\}$, over a prediction horizon of $H=25$ steps, approximately one shedding cycle, that minimises a predefined cost function $J_{\text{MPC}}$, which will be introduced later.

\subsubsection{Cost function}
The cost function is designed to achieve the primary goal of drag reduction while adhering to practical control constraints. The total cost, $J_{\text{MPC}}$, is a weighted sum of terms penalising undesirable behaviour over the prediction horizon:
\begin{equation}
\label{eq:mpc_cost_total}
J_{\text{MPC}} = J_{\text{drag}} + J_{\text{lift}} + J_{\text{control}}
\end{equation}
where the drag, lift, and control components are defined below.

The cost related to drag, $J_{\text{drag}}$, combines two objectives: to minimise the mean drag relative to a baseline and to penalise large fluctuations in drag to encourage stability of the wake.
\begin{equation}
\label{eq:mpc_cost_drag}
J_{\text{drag}} = \underbrace{\left(\frac{1}{H}\sum_{k=0}^{H-1}\hat{C}_{d,t+k}\right) - C_{d,\text{ref}}}_{\text{Mean drag increment}} + \underbrace{w_{\text{amp}}\left(\max_{k}(\hat{C}_{d,t+k}) - \min_{k}(\hat{C}_{d,t+k})\right)}_{\text{Drag fluctuation}}
\end{equation}
Here, $C_{d,\text{ref}}=1.051$ is the mean drag of the uncontrolled case and $w_{\text{amp}}=0.1$ is a weighting factor. The constant $C_{d,\text{ref}}$ does not affect the gradient-based optimisation; it is included to provide a convenient baseline for monitoring drag reduction during deployment. The lift-suppression cost, $J_{\text{lift}}$, penalises the mean absolute lift coefficient with the weighting factor $w_{C_l}=0.01$ to discourage asymmetric wake states:
\begin{equation}
\label{eq:mpc_cost_lift}
J_{\text{lift}} = w_{C_l}\frac{1}{H}\sum_{k=0}^{H-1} |\hat{C}_{l,t+k}|
\end{equation}

Finally, the control cost, $J_{\text{control}}$, regulates the actuation signal. Effective regularisation is critical for two reasons: first, from a practical standpoint, smooth control signals are necessary to reduce actuator fatigue and energy consumption. Second, from a modelling perspective, rapidly changing control actions can degrade the predictive accuracy of the data-driven model by forcing it into out-of-distribution states not well represented in the training data. 

Although the framework allows for penalising various aspects of the control signal, including total effort and rate of change, empirical tuning revealed that a single term promoting the smoothness of the control sequence was sufficient for effective regularisation. The final control cost is therefore defined as:
\begin{equation}
\label{eq:mpc_cost_control}
J_{\text{control}} = w_{\text{smooth}} \frac{1}{H-1}\sum_{k=0}^{H-2}\|\mathbf{a}_{t+k+1} - \mathbf{a}_{t+k}\|^2
\end{equation}
This \textit{control smoothness} term penalises the mean square difference between consecutive actions within the prediction horizon. In our implementation, it is weighted with $w_{\text{smooth}} = 8.0$, which was found to produce stable and effective control actions.

All cost terms operate on unscaled physical quantities: $C_d$ and $C_l$ are of order unity, while the squared action increments are of order $10^{-3}$ given the actuator bounds ($a \in [-0.075, 0.075]$), which motivates the larger value of $w_\mathrm{smooth}$. The weights were selected through iterative manual tuning of the closed-loop performance. Automatic hyperparameter tuning \citep{Marra2024Self} could be used to optimize their values given an auxiliary cost function, although not pursued here.

\subsubsection{Optimisation}

The optimal control sequence, $\{\mathbf{a}^*_{t+k}\}_{k=0}^{H-1}$, is found using a gradient-based optimisation approach. A key advantage of our framework, and a solution to the high computational cost often associated with MPC, is the implementation of the entire predictive model (encoder, dynamics, and decoder) in Pytorch \cite{paszke2019pytorch}. This makes the cost function $J_{\text{MPC}}$ fully differentiable with respect to the sequence of future actions. We can therefore leverage automatic differentiation to compute the exact gradient of the cost function via backpropagation through the unrolled model predictions. This is significantly more efficient than derivative-free methods or numerical approximations, making the optimisation tractable for real-time control.

The optimisation process at each control step is as follows:
\begin{enumerate}
    \item \textbf{Initialisation:} The sequence of actions is initialised. To ensure temporal consistency and accelerate convergence, we employ a warm-start strategy where the initial guess is the optimised sequence from the previous time step, shifted forward by one step.
    \item \textbf{Prediction:} Starting with the current latent state $\mathbf{z}_t = f_{\text{enc}}(\mathbf{S}_t)$, the dynamics model is recursively applied for $H$ steps, using actions from the current sequence, to generate a sequence of future latent states $\{\hat{\mathbf{z}}_{t+1}, \dots, \hat{\mathbf{z}}_{t+H}\}$. The force decoder then maps this sequence to the predicted forces $\{\hat{\mathbf{C}}_{t+1}, \dots, \hat{\mathbf{C}}_{t+H}\}$.
    \item \textbf{Optimisation:} The Adam optimiser \cite{adam2014method} is used to update the action sequence for a small number of iterations (typically 5), using a learning rate of $10^{-3}$, to minimise the cost $J_{\text{MPC}}$. Control actions are constrained to their physical limits at each iteration.
    \item \textbf{Receding horizon:} Following the receding horizon principle, only the first action of the final optimised sequence, $\mathbf{a}^*_t$, is applied to the flow simulation. The rest of the sequence is discarded (except for initialisation in the next step), and the whole process is repeated in the next time step $t+1$, using new sensor measurements.
\end{enumerate}

A sensitivity analysis showed that the solution is largely insensitive to the Adam learning rate across two orders of magnitude ($10^{-4}$ to $10^{-2}$), with the only effect being slow convergence speed for the lower value and increased noise sensitivity for the largest value. Additionally, 20 independent random initialisations converge to tightly clustered cost values at each tested snapshot, indicating that local minima are not problematic for this surrogate-based optimisation landscape.

The training dataset consists of 50000 snapshots, generated over approximately 62 hours on a 96-core Threadripper CPU running OpenFOAM with 32 parallel threads. Model training requires approximately 30 minutes on a NVIDIA L40s GPU. Once trained, the MPC optimisation at each control step takes approximately 100\,ms on the same hardware, without specific inference optimisation.

\section{\label{sec:results}Results}

\subsection{Model performance and latent space structure}
\label{sec:modelPerf}

\begin{figure}[htbp]
    \centering
    \includegraphics[width=1.\textwidth]{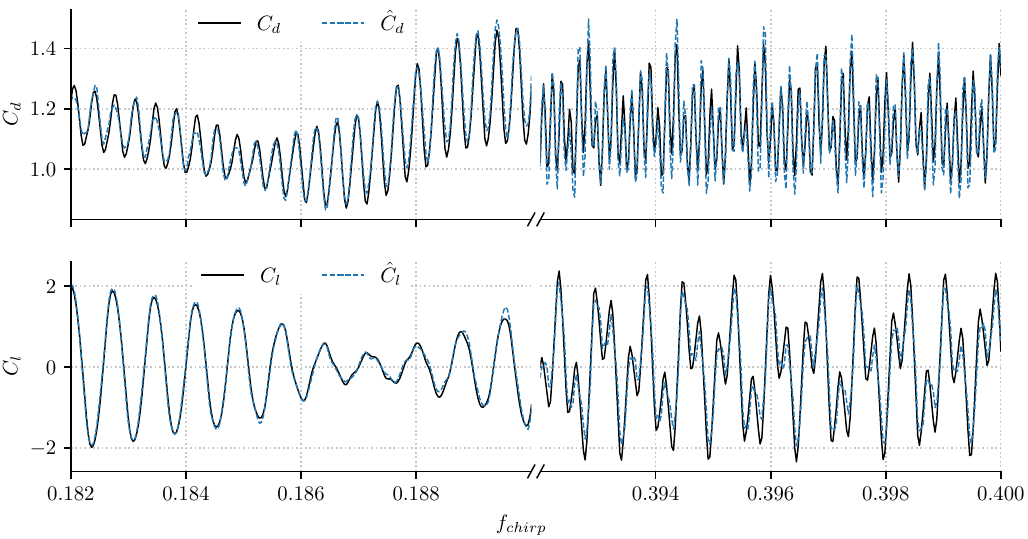}
    \caption{Comparison between the aerodynamic force coefficients estimated by the force decoder and the ground truth values from the test dataset using a chirp actuation signal with instantaneous frequency $f_{chirp}$. The plot shows a range where the wake briefly stabilises and the highest frequency region. The dashed blue line represents the prediction and the solid black line represents the reference values.}
    \label{fig:test_forces}
\end{figure}

\begin{figure}[htbp]
    \centering
    \includegraphics[width=1.\textwidth]{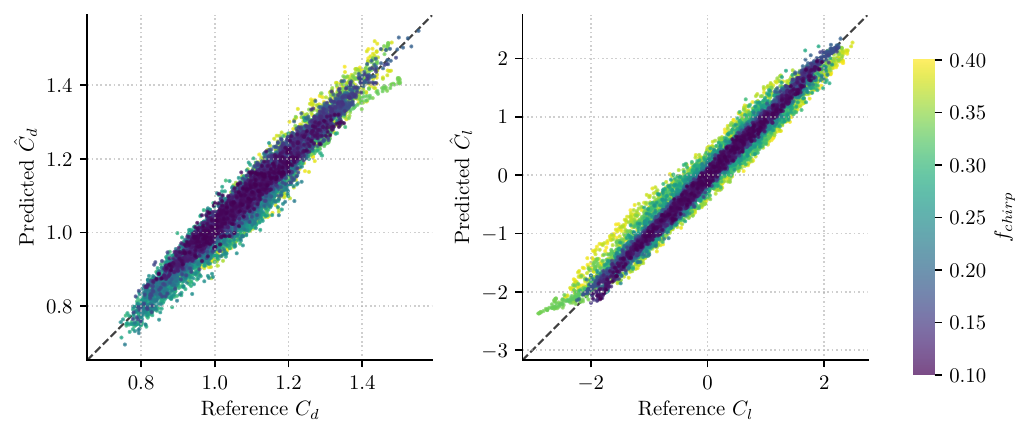}
    \caption{Comparison between the aerodynamic force coefficients estimated by the force decoder and the ground truth values from the test dataset using a chirp actuation signal. The colour scale represents the instantaneous chirp frequency $f_{chirp}$ of the actuation. The model achieves $R^2(C_d)=0.94$ and $R^2(C_l)=0.96$.}
    \label{fig:test_forces_regression}
\end{figure}

The performance of the trained model is evaluated on the test dataset, which was generated using a chirp actuation signal to assess the generalisation capabilities of the model across a range of frequencies. To evaluate the performance of the force decoder, Figure~\ref{fig:test_forces} presents a comparison between the mapped aerodynamic forces and the ground truth values of the dataset, using only the encoder and decoder networks, without prediction. The model demonstrates high fidelity in its estimates, achieving a coefficient of determination ($R^2$) of 0.94 for $C_d$ and 0.96 for $C_l$ and normalised error ($L_1/\sigma$) of 0.19 for $C_d$ and 0.15 for $C_l$ , where the normalised error $L_1/\sigma$ is the mean absolute error normalised by the standard deviation of the reference signal on the test set. This finding verifies that the model accurately predicts instantaneous forces from sensor history across various actuation frequencies, including wake stabilisation areas, thus offering a reliable basis for the control system (see Fig.~\ref{fig:test_forces_regression}).

\begin{figure}[htbp]
    \centering
    \includegraphics[width=1.\textwidth]{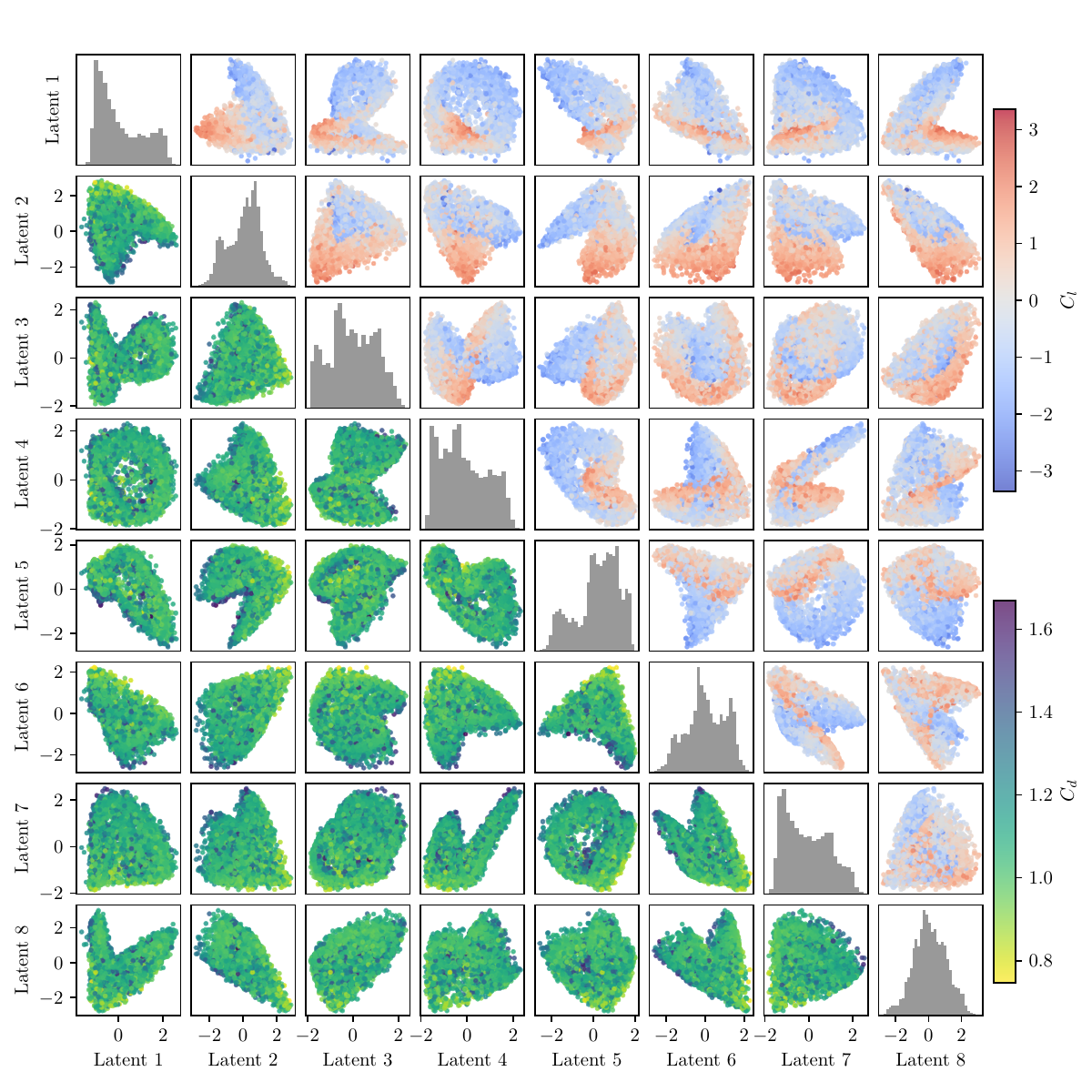}
    \caption{Pairs plot of latent space variables from train data. Diagonal: histogram of each latent variable. Top: Coloured by $C_l$. Bottom: Coloured by $C_d$}
    \label{fig:latent_pairs}
\end{figure}

A key aspect of our framework is the emergence of a low-dimensional latent space that is not only predictable, but also physically meaningful. To investigate its structure, Figure~\ref{fig:latent_pairs} presents a pair plot of the eight latent space variables from the training data. The diagonal elements display histograms for each latent variable, while the off-diagonal plots, coloured by the instantaneous lift (top triangle) and drag (bottom triangle) coefficients, illustrate the relationships between pairs.
The visualisation reveals a highly-structured latent space. For example, the relationship between latent variables 1 and 2 forms a distinct V-shaped manifold. Each arm of the V corresponds to an opposite sign of the lateral force, while the drag force remains symmetric and reaches its minimum at the vertex. This structure is characteristic of the limit-cycle dynamics associated with vortex shedding, where drag fluctuations occur at twice the frequency of lift fluctuations. The smooth variation of the force coefficients along these manifolds confirms that the model has successfully embedded the primary shedding dynamics. Furthermore, the histograms show that the latent variables are well-distributed, preventing modal collapse. This, combined with the structure of the off-diagonal plots, demonstrates that the VICReg regularisation successfully achieved its objective of producing a decorrelated and physically meaningful latent space.

\begin{figure}[htbp]
    \centering
    \includegraphics[width=\textwidth]{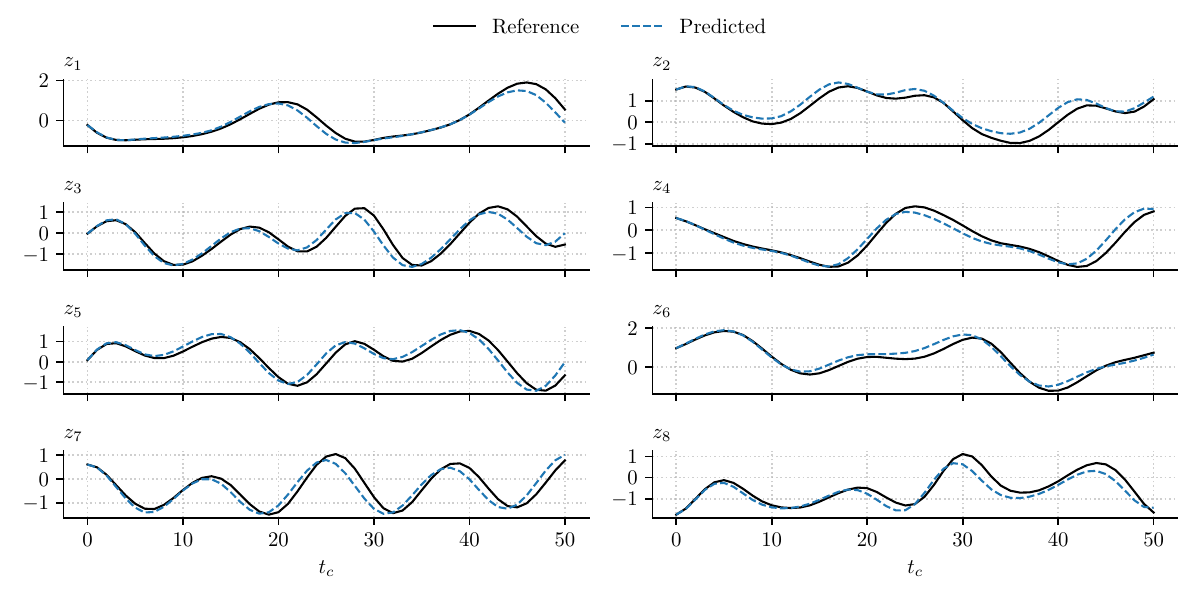}
    \caption{Sample prediction of latent-space variables from test dataset over two shedding cycles. The dashed blue line represents the prediction and the solid black line represents the reference values.}
    \label{fig:latent_prediction}
\end{figure}

The model must also accurately predict the temporal evolution of the system. Figure~\ref{fig:latent_prediction} displays a multistep sample prediction of the latent space variables of the test dataset over two shedding periods. The latent-space dynamics model generates trajectories that closely follow the ground truth obtained from the encoder. This confirms the effectiveness of the recurrent dynamics model in capturing the underlying temporal patterns of the flow, which is a prerequisite for the predictive control task.

\subsection{Efficacy of SHAP-based sensor selection}

\begin{figure}[htbp]
    \centering
    \includegraphics[width=0.9\textwidth]{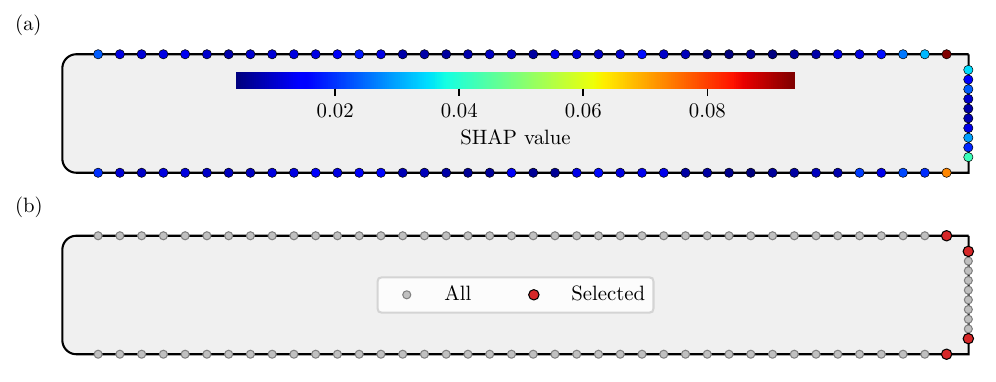}
    \caption{Results of the sensor selection algorithm. (a) Sensor importance SHAP values obtained for each of the sensors. (b) Final selected sensors used in all the results.}
    \label{fig:SHAP}
\end{figure}

An important step towards a practical implementation is minimising the required sensor count. Following training of the latent dynamics model with the full set of $90$ sensors, we employ the SHAP methodology to identify the most informative sensors to predict the state in the latent coordinate space.
The importance of each sensor, quantified as the mean absolute SHAP value, is visualised in Figure~\ref{fig:SHAP} (a). The results clearly indicate that the most critical sensors are concentrated at the base of the vehicle. This finding is physically intuitive, as these sensors are closer to the region affected by flow detachment; thus, they are sensing the pressure fluctuations associated with the vortex shedding dynamics of the wake. In contrast, sensors located further upstream on the sides of the vehicle, where the flow is largely attached, contribute significantly less information to the model, at least as far as the wake control is concerned.  This finding indicates that the SHAP analysis identifies the physically relevant regions for sensing the wake, although the specific sensor selected within each region is subject to the strong local correlation of the surface pressure field. We assess the stability of this selection and its sensitivity to the multicollinearity of the sensor array in \ref{app:shap_robustness}.

\begin{figure}[htbp]
    \centering
    \includegraphics[width=\textwidth]{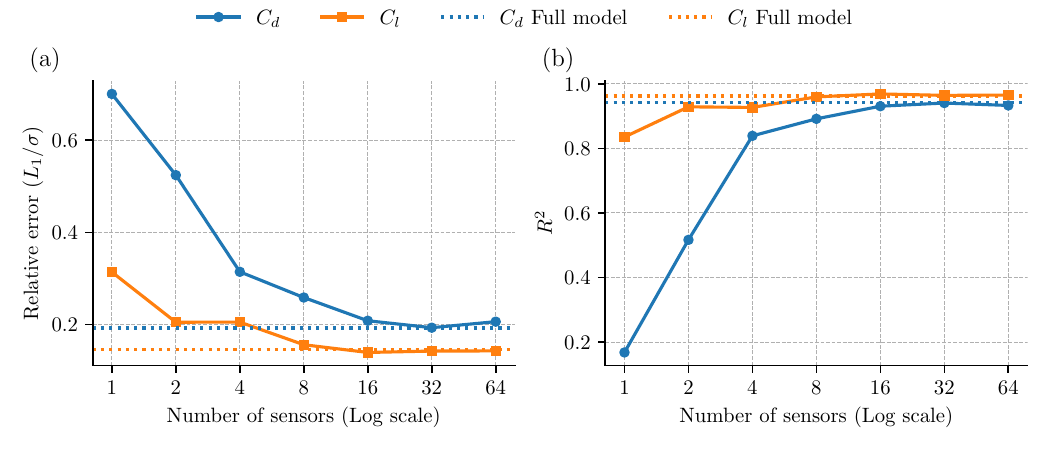}
    \caption{Relative force prediction error (a) and $R^2$ metrics (b) on the test dataset as a function of the number of sensors used. The error is the $L_1/\sigma$ metric for $C_d$ and $C_l$. Sensors are added according to their SHAP-based importance ranking.}
    \label{fig:error_vs_sensors}
\end{figure}

To evaluate the trade-off between the number of sensors and predictive accuracy, a series of ``slim'' encoders were trained using knowledge distillation for progressively smaller subsets of sensors, ranked by their SHAP importance. The performance of each ``slim'' encoder was assessed by pairing it with the original frozen force decoder, evaluating it on the test dataset. Figure~\ref{fig:error_vs_sensors} plots the resulting force prediction error and $R^2$ against the number of sensors used. The error is quantified as the $L_1/\sigma$ metric for both $C_d$ and $C_l$, as defined in Section~\ref{sec:modelPerf}, to provide a relative measure of performance.
The plot shows an initial decrease in $C_d$ error as the first few sensors are added, with the performance quickly plateauing. Using the top 16 sensors achieves a force prediction accuracy nearly identical to that of the entire 90-sensor array. This demonstrates a significant redundancy in the initial sensor set and confirms that a small, strategically placed subset of sensors can capture the essential flow dynamics required for accurate force estimation.

Based on this analysis and considering the diminishing returns of adding more sensors, the top four sensors were selected for the final control implementation, as shown in Figure~\ref{fig:SHAP} (b). This minimal sensor configuration achieves a good balance between high-fidelity state estimation and practical constraints of potential real-world deployment.

\subsection{Closed-loop control performance}
\begin{figure}[htbp]
    \centering
    \includegraphics[width=1.\textwidth]{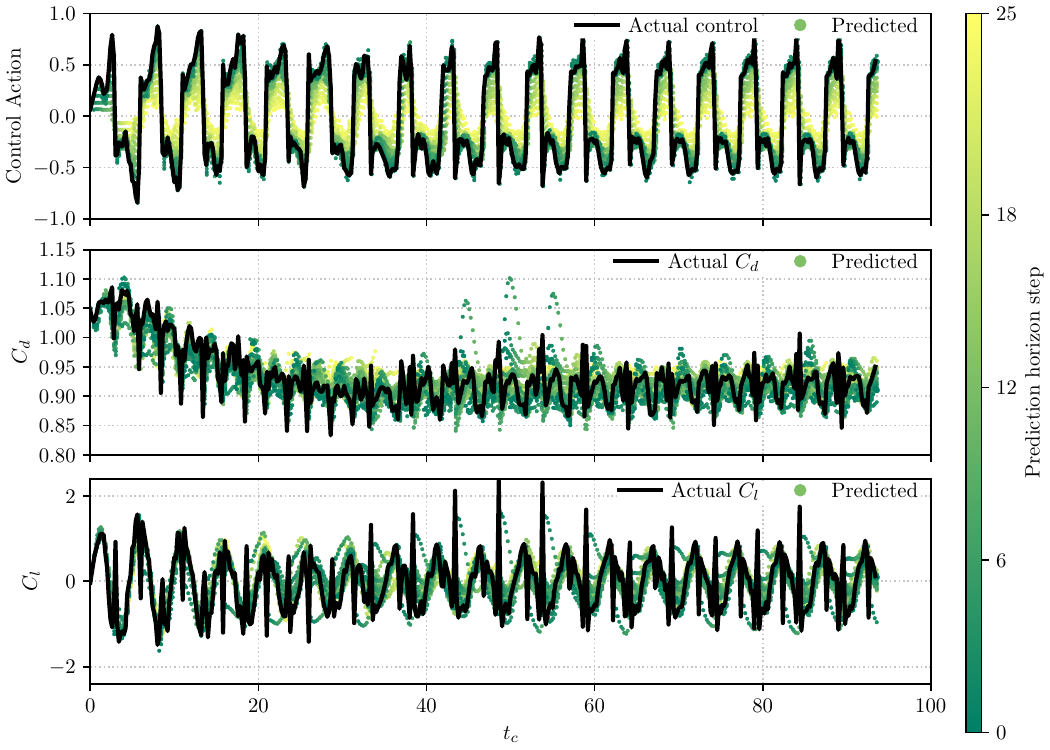}
    \caption{Time series from the closed-loop MPC implementation. Coloured points show the predictions of the model over the optimisation horizon, converging to the final value (solid black line) as the time step is reached. From top to bottom: Normalised control action, $C_d$, and $C_l$.}
    \label{fig:MPC}
\end{figure}

The MPC controller, integrated with the lightweight ``slim'' encoder operating on only four pressure sensors, was deployed in the high-fidelity DNS to assess its performance. The results of the closed-loop control are presented in Figure~\ref{fig:MPC}, which displays the time evolution of the control action, drag, and lift coefficients. The controller successfully reduces the mean drag coefficient by 12.8\%, from a baseline of 1.051 in the uncontrolled case to an average of 0.916 under MPC, a result comparable to related experimental work \cite{pastoor2008dshaped}. For reference, in the chirp test dataset, open-loop periodic forcing at the shedding frequency produces only transient drag reduction ($C_d \approx 1.0$) when the actuation phase briefly synchronises with the wake (visible in Figure~\ref{fig:test_forces}), but drag returns to uncontrolled levels as the phase drifts, highlighting the need for closed-loop predictive control. The control action is smooth and settles into a quasi-periodic pattern with a frequency that effectively counteracts the natural vortex shedding. The figure also illustrates the accuracy of the underlying predictive model during the online control task; the predictions of the model for the next states mostly align with the ground truth obtained from the simulation, although some discrepancies appear as the wake stabilises and the model departs further from its interpolation region. This highlights the robustness of MPC, which operates effectively despite the limitations of the underlying model. Further robustness analyses are provided in the appendices: \ref{app:ood} evaluates the model on actuation frequencies outside the training band, while \ref{app:robustness_closedloop} assesses closed-loop performance under sensor noise or actuator delay. For discrete actuators, the smoothness penalty already drives the control signal towards quasi-binary behaviour, suggesting compatibility with discrete actuation via thresholding.

\begin{figure}[htbp]
    \centering
    \includegraphics[width=1.\textwidth]{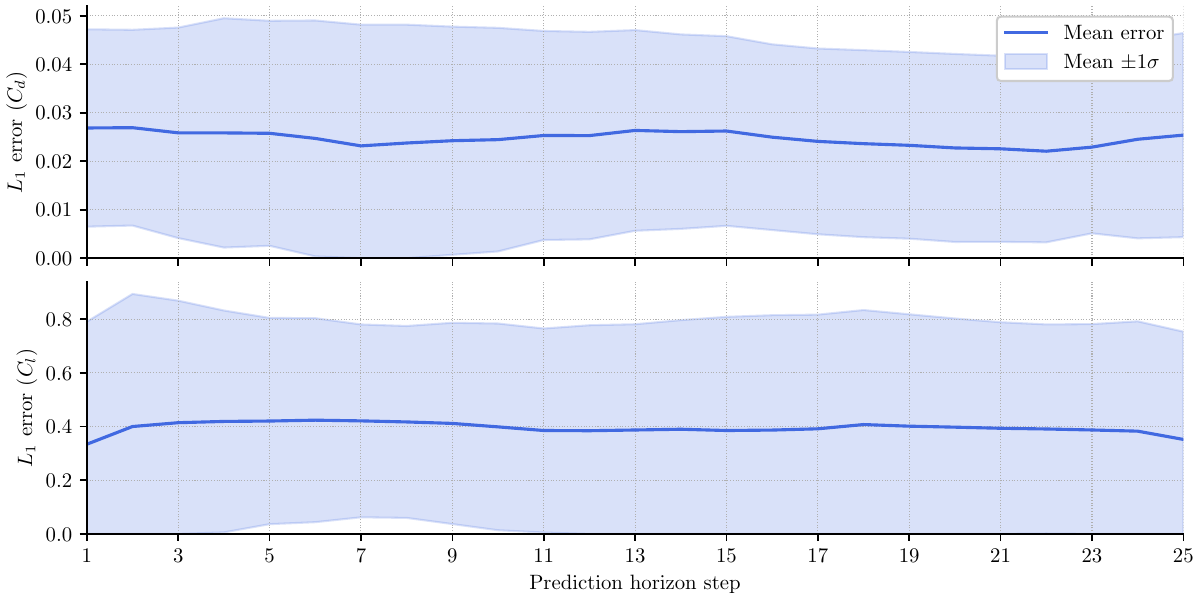}
    \caption{$L_1$ error between the aerodynamic forces estimated by the model during MPC control using four sensors and the final actual values. Solid line represents mean error and shaded area represents the standard deviation of the error. Time increment between steps is $\Delta t = 0.2 t_c$.}
    \label{fig:mpc_forces_error}
\end{figure}

Figure~\ref{fig:mpc_forces_error} shows the error in the prediction of forces along the prediction horizon during MPC control. The error does not show a clear trend over the prediction horizon, which indicates that the error is dominated by encoding and decoding errors instead of the prediction error.

\begin{figure}[htbp]
    \centering
    \includegraphics[width=\textwidth]{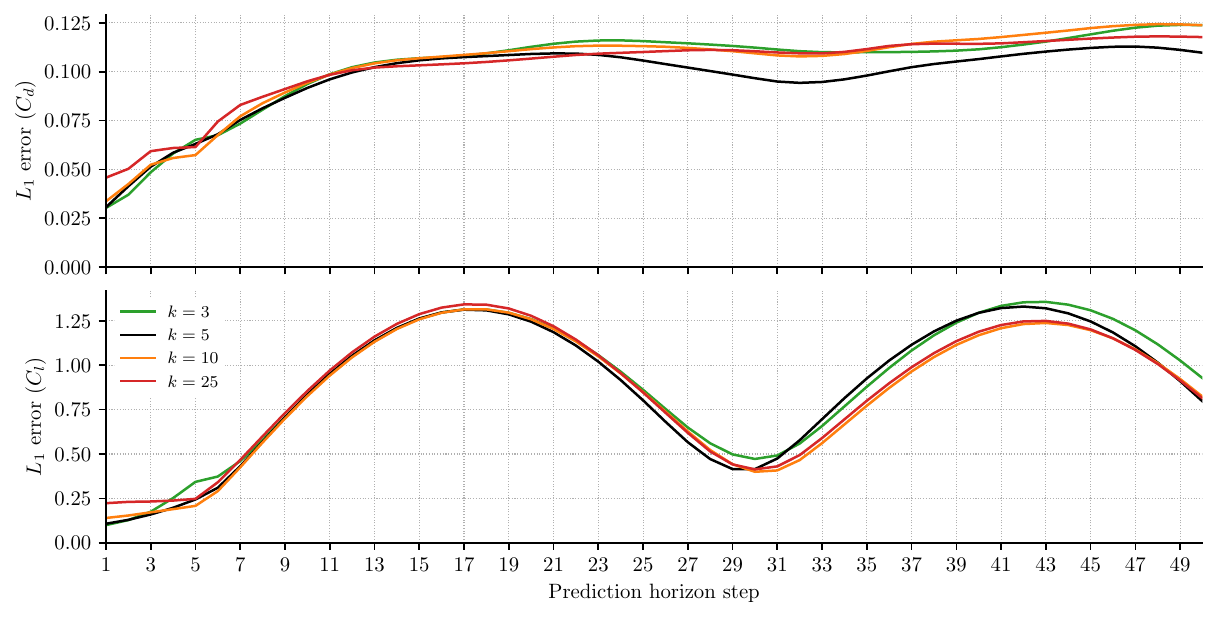}
    \caption{Open-loop $L_1$ averaged force prediction error versus prediction step for models trained with $k \in \{3, 5, 10, 25\}$.  All models produce comparable error growth on the chirp evaluation dataset; the $k=5$ model achieves the lowest steady-state $C_d$ error. Numerical $L_1/\sigma$ values at $h=25$ are reported in the text.}
    \label{fig:horizon_ablation}
\end{figure}

 To assess whether the training unrolling horizon $k=5$ limits accuracy over the MPC prediction horizon ($H=25$), we trained additional models with $k \in \{3, 10, 25\}$ and evaluated all four on open-loop rollouts of 50 steps on the chirp evaluation dataset. All models produce comparable error growth on this dataset; at step $h=25$, the $L_1/\sigma$ errors for $C_d$ are \textbf{0.83}, \textbf{0.76}, \textbf{0.81}, \textbf{0.79} and for $C_l$ are \textbf{0.87}, \textbf{0.81}, \textbf{0.86}, \textbf{0.86}, for $k \in \{3, 5, 10, 25\}$ respectively. The choice $k=5$ achieves the lowest steady-state $C_d$ error while keeping training time to 30 minutes (versus 78 minutes for $k=25$). These results, shown in Figure~\ref{fig:horizon_ablation}, indicate that $k=5$ performs comparably to longer training horizons on this evaluation set, although this does not guarantee long-horizon stability for arbitrary excitations.

\begin{figure}[htbp]
    \centering
    \includegraphics[width=0.7\textwidth]{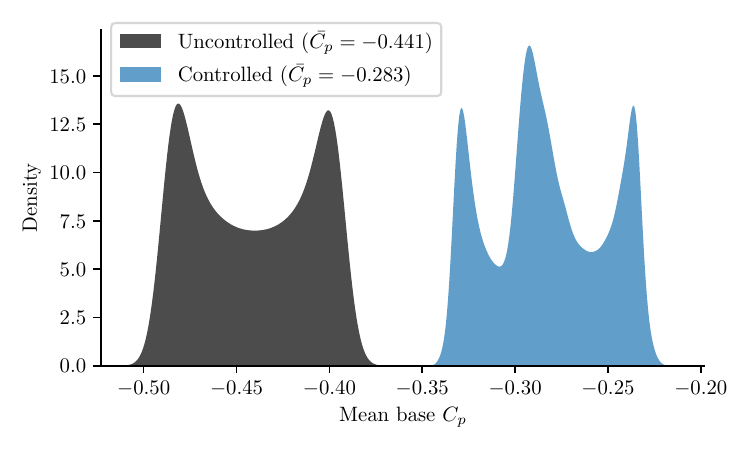}
    \includegraphics[width=0.7\textwidth]{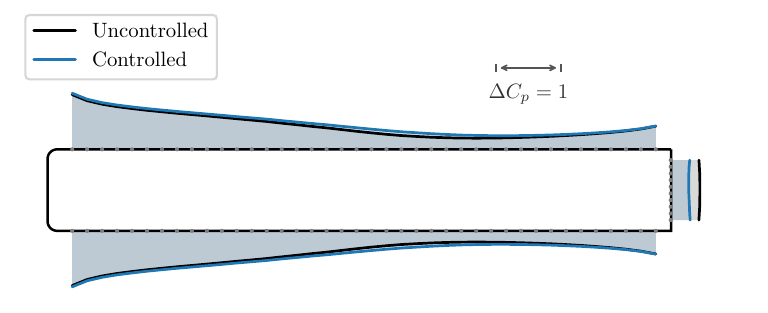}
    \caption{Top: Kernel density estimate (KDE) plot of the mean pressure coefficient ($C_p$) averaged over the sensors at the base of the vehicle. Bottom: Pressure coefficient distribution for the uncontrolled (black) and MPC-controlled (blue) cases, negative coefficient points outwards.}
    \label{fig:back_p_histogram}
\end{figure}

The physical mechanism responsible for this drag reduction is the stabilisation of the wake, achieved by synchronising the jet actuation with the shedding to increase the base pressure.
The MPC significantly weakens the suction at the base of the vehicle, a primary source of pressure drag. This is quantitatively confirmed in Figure~\ref{fig:back_p_histogram}, which shows the distribution of the pressure coefficient ($C_p$) averaged between the base sensors. The control action shifts the mean base $C_p$ from $-0.441$ to a more favourable $-0.283$ for an improvement of more than $35\%$, in line with the results in related problems \cite{dalla2017dshaped}.

\begin{figure}[htbp]
    \centering
    \includegraphics[width=0.8\textwidth]{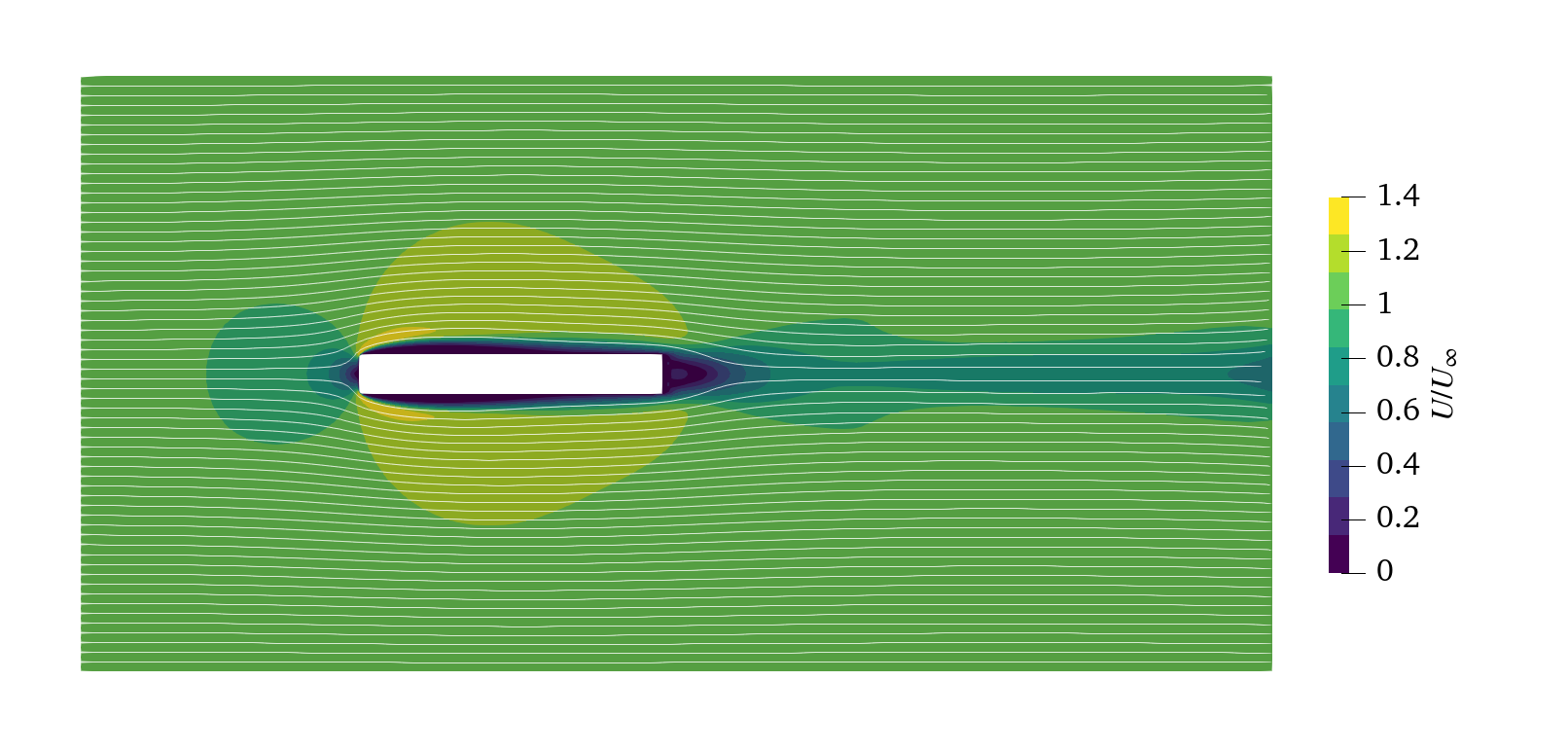}
    \includegraphics[width=0.8\textwidth]{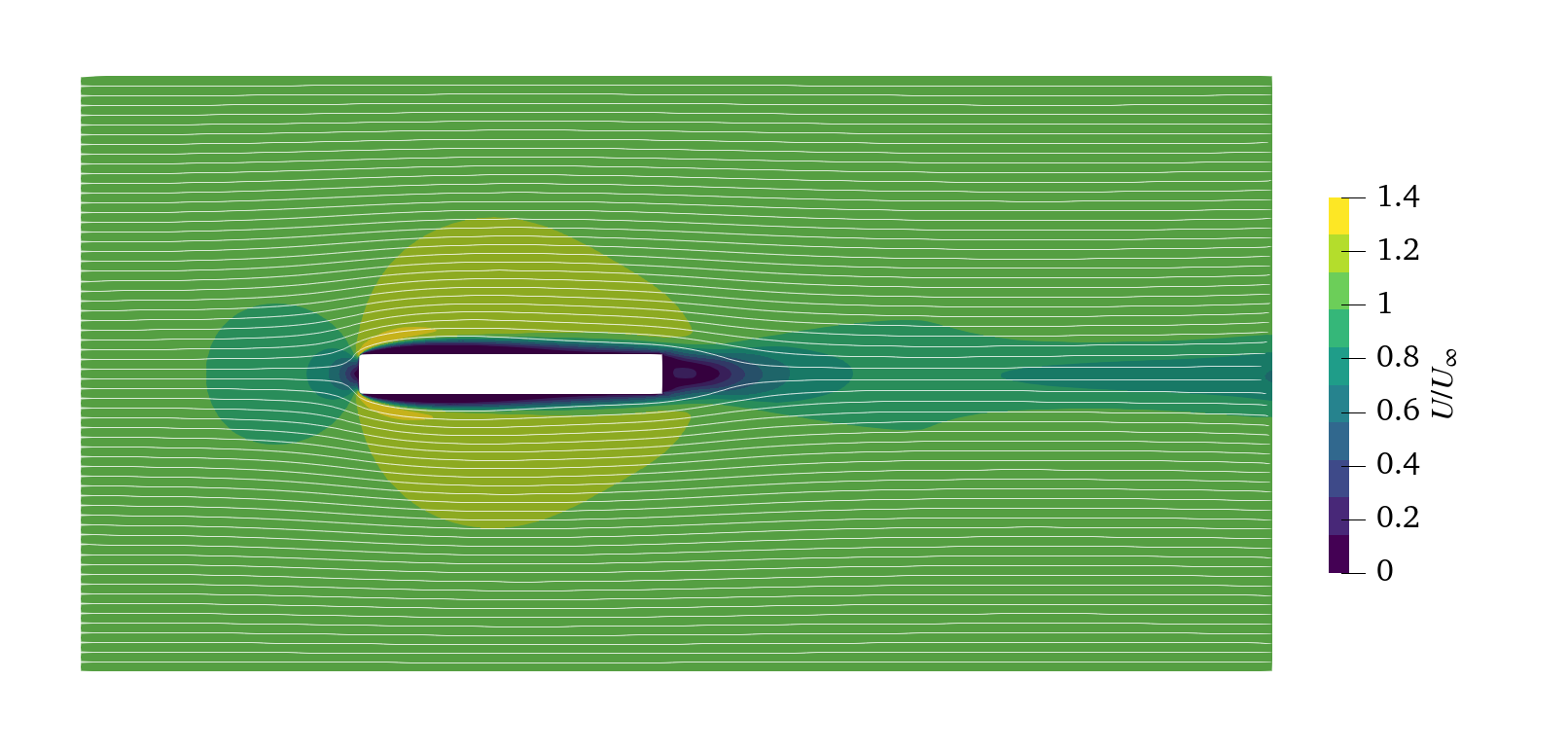}
    \caption{Comparison of the time-averaged streamwise velocity ($U_x$) fields. Top: Uncontrolled baseline case. Bottom: Case with MPC active.}
    \label{fig:Ux_fields}
\end{figure}

\begin{figure}[htbp]
    \centering
    \includegraphics[width=0.7\textwidth]{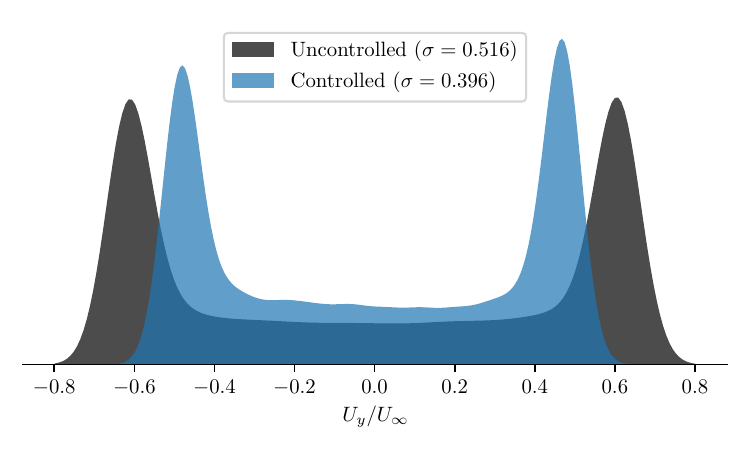}
    \caption{KDE plot of transverse velocity ($U_y$) fluctuations at a probe in the near wake (10, 0), comparing the uncontrolled (black) and MPC-controlled (blue) cases.}
    \label{fig:probe_Uy_histogram}
\end{figure}

This increase in base pressure is a direct consequence of a more stable and organised wake structure. The time-averaged streamwise velocity fields in Figure~\ref{fig:Ux_fields} show that the control extends the recirculation bubble. The suppression of large-scale turbulent fluctuations is further quantified in Figure~\ref{fig:probe_Uy_histogram}, which shows the histogram of transverse velocity fluctuations at a probe located in the near wake at coordinates (10, 0). The distribution for the controlled case is visibly narrower, with the standard deviation of the fluctuations reduced by more than 23\% (from 0.516 to 0.396). This confirms that the MPC strategy achieves drag reduction not through brute force but by efficiently exploiting the dynamics of the wake.

An important observation is that the surrogate model, trained entirely on open-loop data, operates during closed-loop control in a regime not explicitly represented in the training set. The fact that the controller nevertheless maintains stable drag reduction can be understood by examining the nature of the distribution shift. Rather than driving the system into unexplored, high-dimensional regions of the state space, the MPC stabilises the wake into a quasi-periodic regime with reduced vortex shedding that lies close to the training manifold. The smoothness penalty in the MPC cost function reinforces this effect by constraining the rate at which the controller can push the system away from states represented in the training data. The prediction error analysis in Figure~\ref{fig:mpc_forces_error}  provides quantitative support: the error remains flat across the prediction horizon, confirming that model accuracy does not degrade progressively during the controlled episode and that the dominant source of error is the encoder-decoder mapping rather than drift in the latent dynamics.  In the present compact controlled regime, this behaviour suggests that iterative data collection strategies such as DAgger~\cite{ross2011dagger} would be expected to yield diminishing returns after a small number of iterations; however, this remains a hypothesis that has not been formally tested here.

\section{\label{sec:conclusion}Conclusions}

The results presented in this study demonstrate a successful proof-of-concept framework for designing an effective and practical active flow control system. A key distinguishing feature of our approach is the exclusive reliance on non-intrusive, surface-mounted pressure sensors. Many contemporary studies on data-driven flow control use velocity or pressure probes placed directly in the wake to observe the flow state \cite{Rabault2019, Deda2024}. Although this approach is more effective for state estimation, it presents significant challenges for  practical implementation in vehicles, as deploying sensors in the flow field is generally impractical. By constraining our model to use only sensors on the vehicle surface, we establish a path toward a more easily deployable system.

However, this choice presents a more difficult problem. The dynamics of the wake, which is the primary target of control, are not measured directly. Instead, the wake state must be inferred from its pressure footprint on the rear surface of the body, which involves an inherent convective delay and a complex relationship between the shedding structures and the resulting pressure fluctuations. Our methodology successfully addresses this challenge through the temporal encoder. The use of an LSTM network, which processes a history of sensor readings, was crucial for reconstructing a latent state that accurately captures the spatio-temporal dynamics of the wake, effectively learning to capture the time-delayed information.

Another significant contribution of this work lies in the practicality and efficiency of the overall workflow. The latent dynamics model was trained entirely on a pre-computed, open-loop dataset. This approach deliberately decouples the data acquisition and modelling phase from the control design. For physical systems, the primary bottleneck is often not the generation of more data once an experiment is running, but rather the complexity, risk, and implementation overhead of an online learning loop (e.g., for reinforcement learning). Our offline strategy circumvents these issues entirely, providing a robust and straightforward path to deployment: acquire a representative dataset, train the controller offline, and then deploy the static, pre-trained controller. The robustness of the MPC controller, which functioned effectively even as it drove the system into states not well represented in the initial dataset, underscores the feasibility of this model-based approach. This is further enabled by the implementation of the model within a framework that supports automatic differentiation, making the MPC loop computationally efficient enough for real-time application.

The present results should be interpreted within the limitations of the test configuration. The two-dimensional, low-Reynolds-number regime does not capture the three-dimensional complexity, broadband spectral content, or turbulent separation of full-scale ground vehicles ($Re > 10^6$). Extending the framework to realistic conditions requires generating three-dimensional training data at higher Reynolds numbers and a latent dynamics model capable of representing a richer state space. Nevertheless, the physical mechanism exploited by the controller, i.e. base pressure recovery through suppression of coherent vortex shedding, has been demonstrated experimentally at higher $Re$ with the same type of pulsed-jet actuation~\cite{pastoor2008dshaped}, which supports the relevance of the control strategy beyond the present regime. A further limitation is that the offline surrogate is expected to degrade under conditions significantly outside the training envelope, including large Reynolds number changes, actuation amplitudes exceeding the training bounds, or external disturbances that excite unrepresented dynamics. Iterative data collection and domain randomisation are immediate extensions to broaden the model validity at a fixed operating point. A more ambitious and largely open challenge for the community is to develop dynamics models that generalise across Reynolds numbers and geometries, so that the cost of data generation is amortised over families of configurations rather than handled individually for each one. 

Furthermore, the methodology for sensor selection addresses a critical aspect of system design. Although an initial dense array of 90 sensors was used to capture a comprehensive picture of the flow, the SHAP-based analysis provided a systematic and interpretable way to prune this set. The analysis confirmed the physical intuition that the sensors at the base of the vehicle are the most essential to observing the wake, as these sensors are closest to the region where the vortex shedding dynamics starts.  Approximately 16 sensors recover near-full open-loop prediction accuracy (Fig.~\ref{fig:error_vs_sensors}), whereas four sensors are sufficient for the demonstrated closed-loop control task, since the controller requires coverage of the informative regions rather than exhaustive sampling within them. A robustness analysis of the selection (\ref{app:shap_robustness}) shows that SHAP acts here as a region-level selector under multicollinearity: the ranking is reproducible across sample draws, the selected sensors are correlated representatives of four distinct informative regions, and removing any of them increases the prediction error. The subsequent knowledge distillation process allowed us to train a lightweight ``slim'' encoder for this minimal sensor set without retraining the entire dynamics and decoder models, drastically reducing the complexity and recurrent cost of a physical closed-loop control system implementation. In essence, this work presents an end-to-end framework that successfully balances drag reduction with the practical constraints of sensor placement and control implementation.

\section*{Declaration of competing interest}

The authors declare that they have no known competing financial interests or personal relationships that could have appeared to influence the work reported in this paper.

\section*{Data availability}

All datasets and codes used in this work will be made openly available in public repositories upon publication.

\section*{CRediT authorship contribution statement}

\textbf{Alberto Solera-Rico:} Conceptualisation, Methodology, Software, Validation, Formal analysis, Investigation, Data Curation, Writing - Original draft, Visualisation.
\textbf{Carlos Sanmiguel Vila:} Conceptualisation, Methodology, Resources, Writing - Review \& Editing, Supervision.
\textbf{Stefano Discetti:} Conceptualisation, Methodology, Resources, Writing - Review \& Editing, Supervision, Project administration, Funding acquisition.

\section*{Declaration of generative AI in scientific writing}

During the preparation of this work, the authors used Gemini 2.5 and Claude Opus 4.6 in the writing process to improve the readability and language of the manuscript. After using these tools, the authors reviewed and edited the content as needed and assume full responsibility for the content of the published article.

\section*{Funding sources}

This activity is part of the project ACCREDITATION (Grant No TED2021-131453B-I00), funded by MCIN/AEI/ 10.13039/501100011033 and by the “European Union NextGenerationEU/PRTR”.

\section*{Acknowledgements}
The authors are grateful to Dr. Andrea Meilán-Vila for her insightful suggestions and critical review of the manuscript.

\appendix
\section{Model performance outside the training frequency band}
\label{app:ood}

\begin{figure}[htbp]
    \centering
    \includegraphics[width=0.8\textwidth]{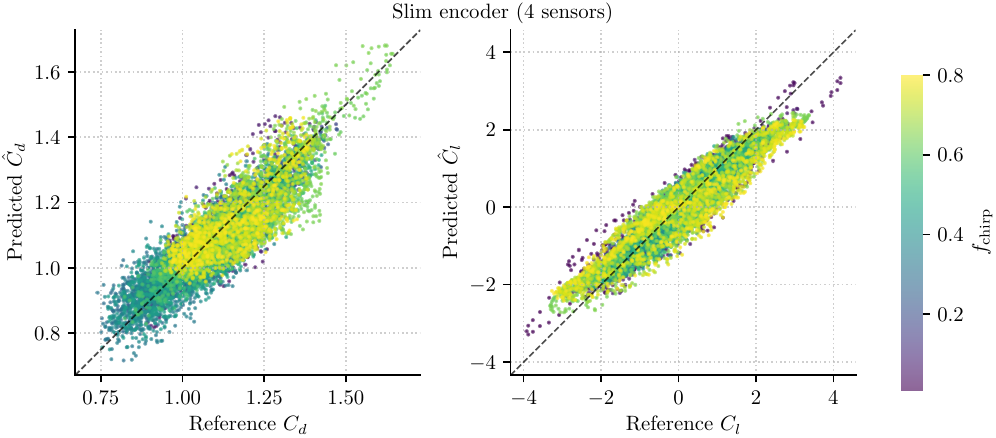}
    \caption{Predicted versus reference force coefficients $C_d$ (left) and $C_l$ (right) for the slim encoder (4 sensors) on the extended chirp dataset ($f \in [0.01, 0.80]$), coloured by chirp frequency.}
    \label{fig:ood_scatter}
\end{figure}

\begin{figure}[htbp]
    \centering
    \includegraphics[width=\textwidth]{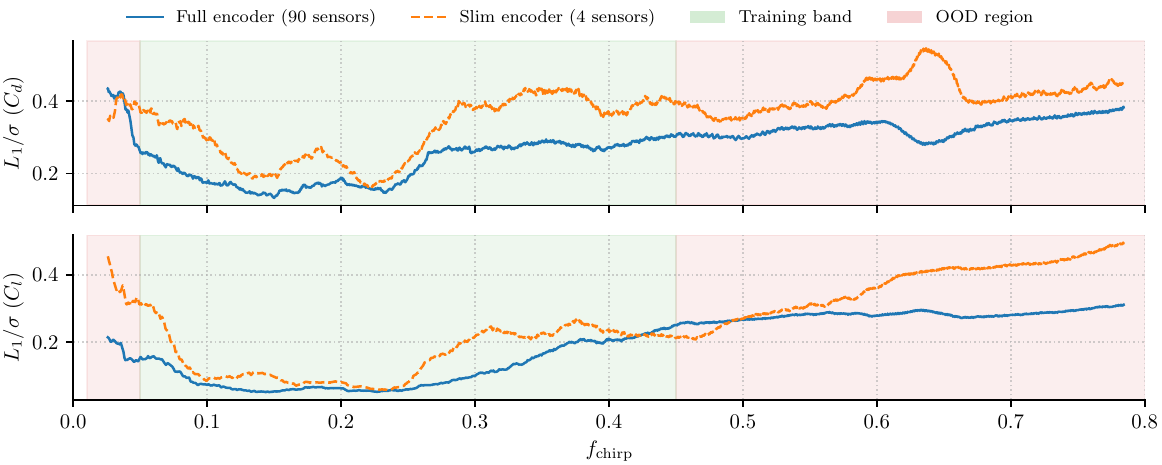}
    \caption{$L_1/\sigma$ prediction error computed over a sliding window of 400 time steps for $C_d$ (top) and $C_l$ (bottom) as a function of chirp frequency. Green: training band ($f \in [0.05, 0.45]$); red: out-of-distribution regions.}
    \label{fig:ood_windowed}
\end{figure}

To assess the model behaviour outside the training frequency band, we generated an extended chirp dataset spanning $f \in [0.01, 0.80]$, beyond the training range ($f \in [0.05, 0.45]$). Figure~\ref{fig:ood_scatter} shows the predicted versus reference force coefficients for the slim encoder (4 sensors) on this dataset. Figure~\ref{fig:ood_windowed} reports the windowed $L_1/\sigma$ prediction error as a function of chirp frequency. Both encoders maintain low error within the training band. Outside this region the error increases but remains bounded, indicating gradual degradation rather than catastrophic failure. The MPC controller operates in the quasi-periodic regime near the shedding frequency, where actuation is most effective for drag reduction; waveforms far from this regime fall outside the intended operating envelope. The evaluation is limited to sinusoidal chirp excitations within the original actuation amplitude envelope; non-chirp waveforms, actuation amplitudes outside the training range, sensor drift or bias, and external disturbances were not tested and remain open directions for future validation.

\section{Closed-loop robustness to sensor noise and actuator delay}
\label{app:robustness_closedloop}

\begin{figure}[htbp]
    \centering
    \includegraphics[width=\textwidth]{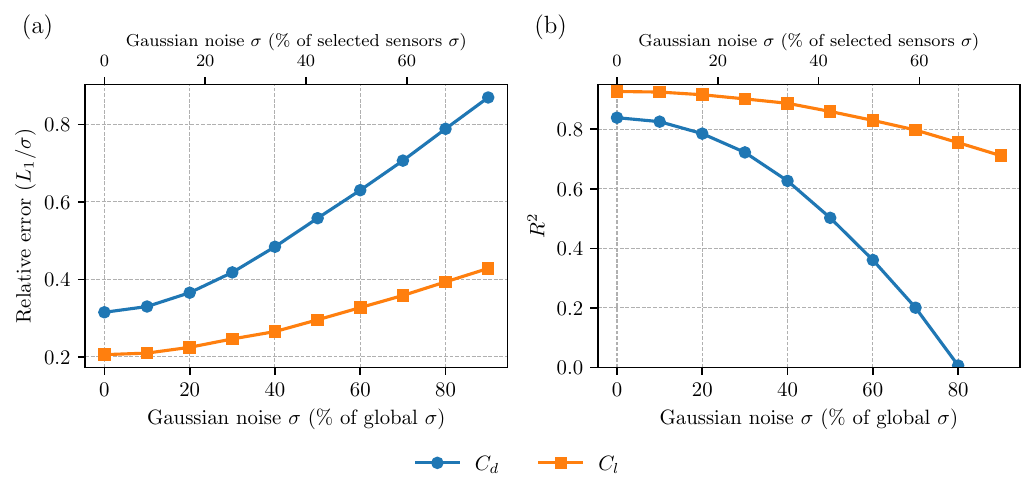}
    \caption{Force prediction robustness of the 4-sensor slim encoder to additive Gaussian noise. (a) Relative $L_1$ error normalised by the force standard deviation and (b) $R^2$, as a function of noise level $\sigma$ expressed as a percentage of the global sensor standard deviation.}
    \label{fig:noise_prediction}
\end{figure}

\begin{figure}[htbp]
    \centering
    \includegraphics[width=\textwidth]{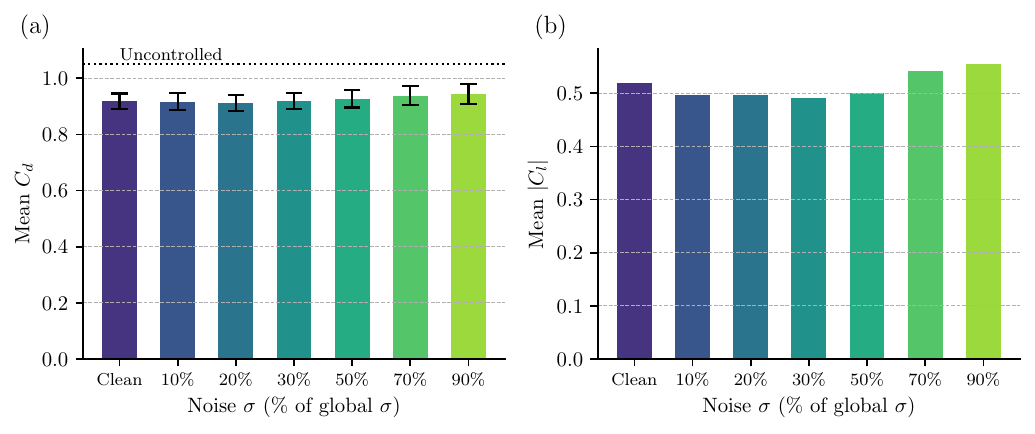}
    \caption{Closed-loop MPC performance under sensor noise. (a) Time-averaged $\overline{C_d}$ and (b) mean $\overline{|C_l|}$ for noise levels from clean to $\sigma = 90\%$ of the global sensor standard deviation. The dotted line indicates the uncontrolled baseline ($C_{d,\mathrm{ref}} = 1.051$). Error bars denote $\pm 1\sigma$ over the controlled episode.}
    \label{fig:noise_mpc}
\end{figure}

To evaluate robustness to sensor corruption, additive Gaussian noise with standard deviation $\sigma$ ranging from 10\% to 90\% of the global sensor standard deviation was applied to the slim encoder inputs on the chirp test dataset (note that noise is smaller in relation to the standard deviation of the selected sensors, as indicated in the secondary horizontal axis in Figure~\ref{fig:noise_prediction}). Figure~\ref{fig:noise_prediction} shows that both $L_1/\sigma$ error and $R^2$ degrade gradually, with $C_l$ remaining well predicted ($R^2 > 0.7$) up to $\sigma = 90\%$, while $C_d$ is more sensitive due to its smaller dynamic range. We attribute this tolerance to the temporal encoder processing a history of 32 time steps, which enables implicit filtering of uncorrelated noise. To verify that this translates to closed-loop performance, full MPC-on-CFD simulations were run under the same noise levels. Figure~\ref{fig:noise_mpc} confirms that the controller maintains consistent drag reduction across all tested noise levels.

\begin{figure}[htbp]
    \centering
    \includegraphics[width=\textwidth]{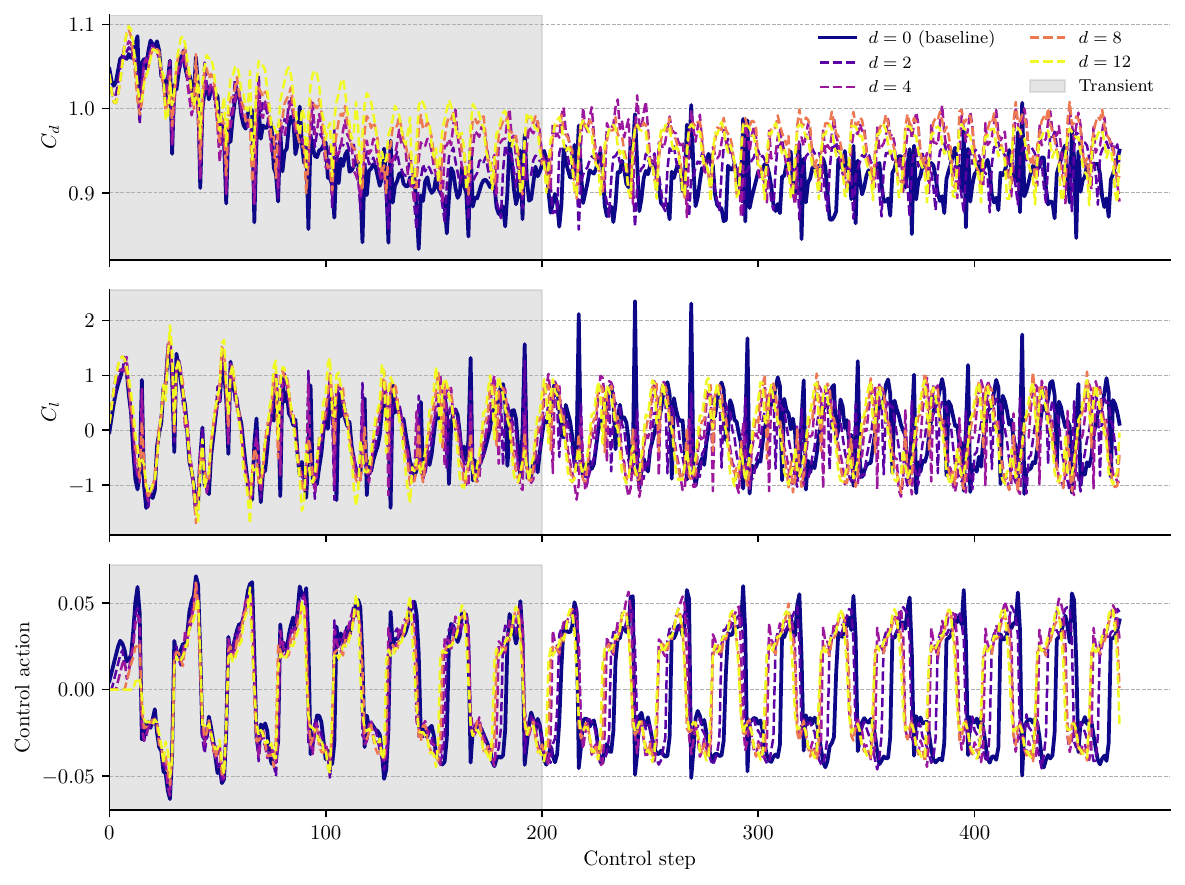}
    \caption{Closed-loop MPC performance under actuator delay. Time evolution of $C_d$ (top), $C_l$ (middle) and control action (bottom) for delays $d \in \{0, 2, 4, 8, 12\}$ control steps ($\Delta t = 0.2\,t_c$). The shaded region denotes the initial transient excluded from steady-state statistics.}
    \label{fig:delay_timeseries}
\end{figure}

\begin{figure}[htbp]
    \centering
    \includegraphics[width=\textwidth]{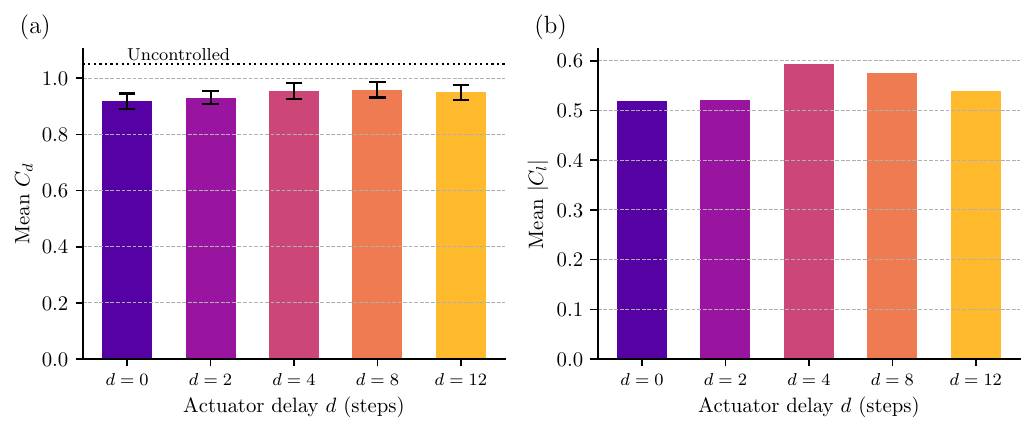}
    \caption{Steady-state aerodynamic coefficients under actuator delay: (a) mean $C_d$ with $\pm 1\sigma$ bars and (b) mean $|C_l|$. The dashed line marks the uncontrolled baseline ($C_d = 1.051$).}
    \label{fig:delay_summary}
\end{figure}

To assess the effect of actuator lag, we model it as a pure input delay of $d$ control steps within the MPC framework. The first $d$ actions in the prediction horizon are frozen to values committed in previous MPC calls but not yet applied, while only the remaining $H - d$ actions are optimised. This does not require retraining the latent dynamics model. Figure~\ref{fig:delay_timeseries} shows the time evolution of $C_d$, $C_l$ and the control action for $d \in \{0, 2, 4, 8, 12\}$. Figure~\ref{fig:delay_summary} summarises the steady-state performance: the controller maintains meaningful drag reduction across all tested delays, with a gradual increase from $\overline{C_d} = 0.918$ ($d = 0$) to $\overline{C_d} = 0.950$ ($d = 12$).

\section{Robustness of the SHAP-based sensor selection}\label{app:shap_robustness}

\begin{figure}[htbp]
    \centering
    \includegraphics[width=\textwidth]{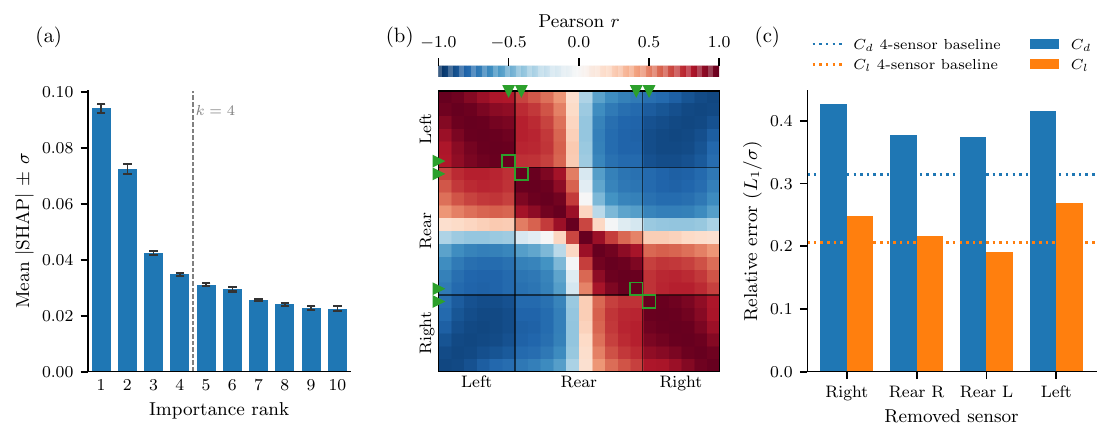}
    \caption{Robustness analysis of the SHAP-based sensor selection. (a) Mean $|\mathrm{SHAP}|$ score $\pm$ one standard deviation over $N=10$ independent \texttt{GradientExplainer} runs with different randomly drawn background ($n_\mathrm{bg}=100$) and explanation ($n_\mathrm{ex}=500$) sample sets; the top-$k$ sensor sets ($k \in \{2,4,6,8\}$) are identical across all ten runs. (b) Pearson correlation matrix around the four SHAP-selected sensors (green markers), grouped as left side, rear face, and right side; the strong within-region correlation and across-region anti-correlation is characteristic of the multicollinear surface pressure field. (c) Leave-one-out relative $L_1/\sigma$ error on the chirp test dataset when each of the four sensors is individually removed and a 3-sensor slim encoder is retrained; dotted lines denote the 4-sensor baseline.}
    \label{fig:shap_robustness_composite}
\end{figure}

The four-sensor configuration used throughout the closed-loop results is obtained by ranking the $90$ surface pressure sensors according to their mean absolute SHAP value, as described in Section~\ref{sec:methodology}. Because surface pressure sensors are spatially dense, the input features are strongly correlated, and SHAP attributions can redistribute importance among correlated variables~\cite{aas2021explaining}. To verify that the selection is not an artefact of this multicollinearity, we report three complementary checks in Figure~\ref{fig:shap_robustness_composite}.

Panel (a) shows the ranking stability across random sample draws. The GradientExplainer analysis was repeated ten times with independently drawn background ($n_\mathrm{bg}=100$) and explanation ($n_\mathrm{ex}=500$) sample sets, while keeping the encoder weights fixed. The standard deviation of the mean $|\mathrm{SHAP}|$ score is negligible compared to its mean, and the top-$k$ sensor sets ($k \in \{2,4,6,8\}$) are identical across all ten runs.

Panel (b) shows the Pearson correlation matrix around the four selected sensors, grouped into the three informative regions (left trailing edge, rear face, right trailing edge). Sensors on the same side are strongly correlated ($r \approx 0.73$ within each pair), and sensors on opposite sides are strongly anti-correlated ($|r|$ up to $0.87$). This is the regime in which SHAP attributions are known to be unstable in terms of which specific feature is credited~\cite{aas2021explaining}.

Panel (c) shows a leave-one-out retraining experiment. For each of the four selected sensors, a 3-sensor slim encoder was retrained without that sensor, and the relative $L_1/\sigma$ error on the chirp test dataset is reported. Removing any of the four sensors increases or matches the prediction error relative to the 4-sensor baseline, with the largest impact on $C_d$. No single sensor is fully redundant, indicating that the algorithm is not simply selecting four near-duplicates of one informative location.

Taken together, the three panels support a region-level interpretation of the SHAP selection in this configuration. The algorithm consistently identifies four spatially distinct informative regions and selects one representative within each; any sensor strongly correlated with one of these would likely yield comparable performance, while removing any of the four increases the prediction error. This is consistent with the diminishing-returns trend in Figure~\ref{fig:error_vs_sensors}: four sensors suffice for the closed-loop control task because it requires coverage of the informative regions rather than exhaustive sampling within them, whereas recovering near-full open-loop prediction accuracy requires approximately 16 sensors.

\bibliographystyle{elsarticle-num}
\bibliography{bibliography}

\end{document}